\documentclass[longbibliography, amssymb, amsmath, aps, prr, reprint, superscriptaddress]{revtex4-2}
\usepackage[caption=false,subrefformat=parens,labelformat=parens]{subfig}
\usepackage[english]{babel}
\usepackage[utf8]{inputenc}
\usepackage[dvipsnames]{xcolor}
\usepackage[colorinlistoftodos, color=green!40, prependcaption]{todonotes}
\usepackage{amsthm}
\usepackage{mathtools}
\usepackage{physics}
\usepackage{gensymb}
\usepackage{xcolor}
\usepackage{subfig}
\usepackage{graphicx}
\usepackage{adjustbox}
\usepackage[T1]{fontenc}
\usepackage{lipsum}
\usepackage{csquotes}
\usepackage[pdftitle={Article}, pdfauthor={Author}]{hyperref} 
\usepackage[capitalise]{cleveref}
\usepackage{titlesec}
\titlespacing{\section}{0pt}{1eM}{0.8em}
\titlespacing{\subsection}{0pt}{0.8em}{0.4em}

\begin{document}

\title{Stick-slip in a stack: how slip dissonance reveals aging}

\author{Samuel Poincloux}
\affiliation{École Polytechnique Fédérale de Lausanne (EPFL), Flexible Structures Laboratory, CH-1015 Lausanne, Switzerland}
\affiliation{Department of Physics, The University of Tokyo, Tokyo, Japan}

\author{Pedro M.~Reis}
\affiliation{École Polytechnique Fédérale de Lausanne (EPFL), Flexible Structures Laboratory, CH-1015 Lausanne, Switzerland}

\author{Tom W.J.~de Geus}
\email[Correspondence email address:]{tom@geus.me}
\affiliation{Institute of Physics, École Polytechnique Fédérale de Lausanne (EPFL), CH-1015 Lausanne, Switzerland}

\begin{abstract}
    We perform physical and numerical experiments to study the stick-slip response of a stack of slabs in contact through dry frictional interfaces driven in quasistatic shear.
    The ratio between the drive's stiffness and the slab's shear stiffness controls the presence or absence of slip synchronization.
    A sufficiently high stiffness ratio leads to synchronization, comprising periodic slip events in which all interfaces slip simultaneously.
    A lower stiffness ratio leads to asynchronous slips and, experimentally, to the stick-slip amplitude becoming broadly distributed as the number of layers in the stack increases.
    We interpret this broadening in light of the combined effect of complex loading paths due to the asynchronous slips and creep.
    Consequently, the aging rate of the interfaces can be readily extracted from the stick-slip cycles, and it is found to be of the same order of magnitude as existing experimental results on a similar material.
    Finally, we discuss the emergence of slow slips and an increase in aging-rate variations when more slabs are added to the stack.
\end{abstract}


\maketitle

\section{Introduction}

Multiple frictional interfaces coupled by elasticity are ubiquitous in everyday objects including books~\cite{Alarcon2016,Poincloux2021}, textiles~\cite{Poincloux2018,Warren2018,Seguin2022}, and multilayer composites~\cite{Talbott1990,Scherer1991}.
In geology, systems comprising multiple frictional interfaces are the norm rather than an exception.
For example, layered rocks such as shale can show multiple sliding interfaces under shear~\cite{Sarout2017,Zuo2020}.
At the larger scales relevant for terrestrial faults, slips producing earthquakes are usually not isolated but embedded into complex fault networks~\cite{Scholz2019}.
The mechanical response of such assemblies of frictional interfaces involves the coupling between the elastic deformation of the layers and the barriers to sliding of the interfaces.

Predicting the onset of slipping is a long-standing problem, even for a single frictional interface~\cite{Popova2015}.
Physical insight and understanding of this class of problems have been driven primarily by high-precision experiments of sliding PMMA blocks whose optical transparency enabled the spatiotemporal tracking of the local contact area~\cite{Ben-David2011}.
These pioneering experiments have elucidated that the onset of slip involves a rupture front that `unzips' the interface.
A correlation with strain measurements close to the interface showed that the stress field and dynamics of the front are well described by fracture mechanics with the fracture energy as the sole fitting parameter~\cite{Svetlizky2014}.
However, the mechanism underlying the nucleation of the rupture front remains elusive, primarily due to experimental limitations, for which novel protocols are being proposed~\cite{Gvirtzman2021}.

From a theoretical perspective, the most common models for the onset of frictional slippage~\cite{Ruina1983,Rice1983,Dieterich1972,Dieterich1979} capture the phenomenology that sliding starts, and after a transient, continues in a steady state when the shear force $F$ balances the friction forces $\mu N$, where $N$ is the normal force and $\mu$ is the ``friction coefficient''.
In these ``rate-and-state'' models, $\mu$ depends nonlinearly on the slip rate $v$ and history $\theta$: $\mu = \mu(v, \theta)$.
At intermediate values of $v$, the friction coefficient is usually assumed to display slip weakening ($\mu$ is a decreasing function of $v$) such that the interface is unstable.
During slip nucleation, the elasticity and inertia of the bulk have a stabilizing effect~\cite{Zheng1998,Bar-Sinai2019,Brener2018,deGeus2022}, such that there exists an effective flow curve whose steady-state displays a minimum $\mu_c$ at $v = v_c$~\cite{Zheng1998}.
Consequently, any perturbation decays and vanishes if the applied load is $F / N < \mu_c$~\cite{BarSinai2012}.
At higher applied loads, the interface destabilizes if a slip patch reaches a critical size \cite{Brener2018} beyond which its dynamics are well described by a sharp rupture front~\cite{Zheng1998,Brener2018,Barras2019} that can be modeled by linear elastic fracture mechanics~\cite{Svetlizky2014,Svetlizky2017b,Svetlizky2020}.
A significantly debated question is the nature of the instability and consequently, how the critical size diverges as a function of $F / N - \mu_c$, e.g.~\cite{Brener2018,Ben-David2011,deGeus2022,Albertini2021}.

A direct consequence of the phenomenology described above is that the interface can display \emph{stick-slip} behavior when driven quasistatically (at a rate $V \ll v_c$).
The link between the stick-slip amplitude and the parameters of the rate-and-state models is debated~\cite{Schar2020,Brener2018,deGeus2022,deGeus2019,Castellano2022,Gvirtzman2021,Bar-Sinai2013}.
One of the authors \cite{deGeus2019,deGeus2022} recently proposed an encompassing theory linking the cited rate dependence and the nucleation of an instability by avalanches of microscopic failures.
Beyond this athermal view, it is a well-known experimental fact that the initial onset to sliding is history-dependent and increases with the time that the interfaces were at rest~\cite{Heslot1994,Baumberger2006,Dokos1946,Dieterich1972}.
This aging behavior is associated with creep~\cite{Dieterich1994} and described by rate-and-state models where the variable $\theta$ introduced above is regarded as time.
Creeping of the interfaces must affect the stick-slip amplitude, but disentangling its contribution is challenging because slip events occur at a narrowly distributed interval (on which statistical fluctuations of the stick-slip amplitude overwhelm the effect of aging \cite{Ben-David2011}).

Beyond a single frictional interface, when multiple frictional interfaces are present, the elasticity of the bulk may potentially lead to a non-trivial coupling.
For example, elastic interactions between faults may strongly affect their slip dynamics~\cite{Harris1991,Romanet2018}.
In addition, acoustic waves transmitted through the elastic bulk may lead to remote triggering of earthquakes~\cite{Johnson2008,Hill1993}, though the large temporal separation suggests a complex coupling.
Predicting the mechanical response of an assembly of elastic frictional interfaces is then a formidable but important challenge.
In particular, identifying the key parameters coupling the layers together and elucidating the role of the number of interfaces are open questions.

Here, we report results from a combined experimental and numerical investigation on the quasistatic stick-slip response of a stack of elastic slabs in contact through frictional interfaces.
Based on the ratio between the stiffness of the drive and the shear stiffness of the slab, we distinguish two regimes: ``stiff'' and ``compliant'' driving.
In the stiff-driving regime, our numerical results exhibit periodic slips involving all the layers leading to narrowly distributed force drops.
By contrast, in the compliant-driving regime, we observe both numerically and experimentally a decoupling of the slip events along the different layers, with interfaces sliding one by one.
In the experiments, we find that this loss of periodicity is accompanied by a broadening of the distribution of the stick-slip amplitudes with the number of layers.
We highlight the role of interface creep in this broadening, exposed by the complex loading paths of the interfaces induced by mechanical coupling between the increasing number of layers.
Overall, the stick-slip response of a stack to shear is controlled by the stiffness ratio between the drive and the layers whereby in the stiff-driving case the stack acts as one layer with periodic slips, while in the compliant-driving case a rich coupling between the layers makes slips much more unpredictable.

\section{Definition of the problem}

\begin{figure}[htp]
    \subfloat{\label{fig:1a}}
    \subfloat{\label{fig:1b}}
    \centering
    \includegraphics[width=\columnwidth]{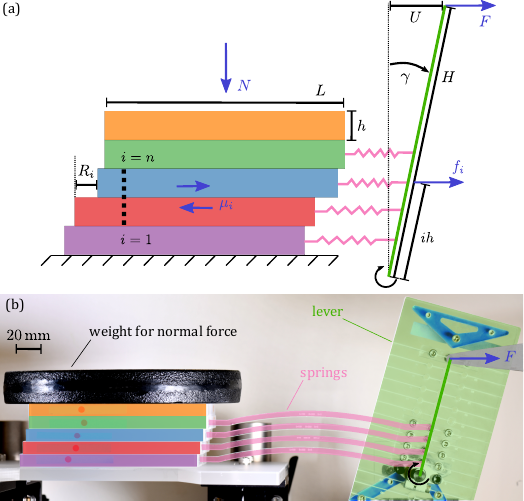}
    \caption{
        (a) Schematic of our model system, shown here with $n = 4$ active (driven) layers.
        The color code of the layers is used throughout the figures.
        (b) Photograph of the corresponding experimental apparatus.
    }
    \label{fig:1}
\end{figure}

We assess the shear response of a model system comprising a stack of $n\in[1,\,5]$ identical slabs of thickness $h$ resting on a surface whose position is fixed.
In \cref{fig:1}, we present a schematic diagram of the system and a photograph of our experimental setup.
Each slab, and its lowermost frictional interface, are numbered as $i = 1, 2, \ldots, n$ from below.
We impose homogeneous shear between all the slabs by connecting each slab through identical springs of stiffness $K$ to a lever that is driven to rotate around a fixed axis at a set rate (\cref{fig:1a}).
The spring connecting to the $i$-th layer is attached to the lever at a distance $ih$ from the rotation axis.

For the drive, we impose the lever's top horizontal displacement $U(t) = V t$, where $t$ represents time, and $V$ is the imposed velocity, taken to be small enough for the interfaces to display stick-slip (such that $V < v_c$~\cite{Baumberger2006,deGeus2022} has to hold).
Thus, our drive imposes a shear rate $\dot \gamma \equiv V / H$, where $H$ is the height of the lever, driving each spring $i$ at a velocity $v_i = i h \dot \gamma$.
During the periods in which the interfaces are `stuck', this drive causes a monotonically increasing shear stress at each of the interfaces.

Our study seeks to address the following questions: (1) What are the relevant parameters controlling the slip synchronization of the interfaces?
(2) How does the shear response evolve with an increasing number of layers $n$?

\section{Stiff vs.~Compliant Driving}

For a system with a single frictional interface, the stick-slip instability occurs only if the driving stiffness $K < K_c$, where $K_c$ depends on the flow properties of the interface and the applied driving rate~\cite{Cochard2003,Baumberger2006,Ruina1983,Rice1983}; a summary of the calculation for a rate-and-state model is provided in Ref~\cite{Baumberger2006}.
The experimental and numerical systems are taken in the stick-slip regime (such that $K < K_c$ has to hold).
We will argue below that with multiple interfaces the rigidity of the drive also controls the degree of slip synchronization.

To gain insight into the effect of the drive on the shear response of a stack, we regard the driving `springs' as a parabolic potential energy imposing the mean position of each slab, such that the slab is free to build up shear.
We now discuss what happens in the limit of stiff and compliant driving, defined next.

\begin{figure}[htp]
    \centering
    \includegraphics[width=\columnwidth]{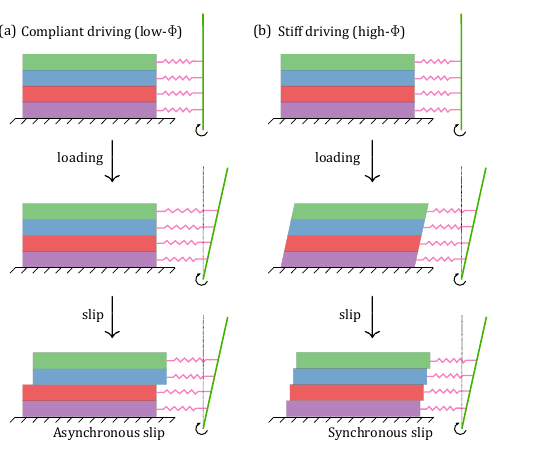}
    \caption{
        Schematic of the loading and consecutive slip of the interfaces in the compliant (a) and stiff (b) regimes.
    }
    \label{fig:rigidity}
\end{figure}

\textit{Rigidity ratio $\Phi$.}
We define the rigidity ratio $\Phi \equiv K / K_s$, where $K$ is the rigidity of the driving springs, and $K_s = A G / h$ is the shear rigidity of the slabs, with $G$ the shear modulus, $A$ the surface area of the frictional interface, and $h$ the slab's height.
This ratio $\Phi$ then quantifies the relative deformation of the driving springs in comparison to the shear deformation of the slabs.
Below, we investigate and discuss two limit regimes: stiff (high-$\Phi$) and compliant (low-$\Phi$) driving.
In the first, significant deformations occur within the layers (and the springs are stiff), whereas, in the second, the deformations occur in the springs (and the layers are stiff).
The loading and slip behaviors of the stacks in these two limit regimes are schematically represented in \cref{fig:rigidity}.

\textit{Stiff driving (high-$\Phi$).}
Let us suppose that the interface $a = 1$ starts slipping.
As the mean position of slab $a$ is fixed by the stiff spring, the slab can only react with a negative shear deformation.
Consequently, the shear stress on the interface above, $i = a + 1$, is increased.
This can trigger a slip on that interface, which in turn causes a stress increase and slip on the interface above for the same reason, and so on for the rest of the stack.
The cascade results in a multi-slip event that erases all memory of the system.
In this case, the multi-layered system thus acts as a system with a single interface with effective properties \footnote{
    Similar considerations hold for other physical systems.
    For example, the statistics of avalanches in spring-block models that are rigidly driven (using overdamped dynamics) differ from those in the case of compliant driving~\cite{Rosso2009,Rosso2007}.
}, showing periodic stick-slip cycles.

\textit{Compliant driving (low-$\Phi$).}
With soft springs, the system can respond to a slip at interface $a$ by advancing the mean position of slabs $i > a$ and compressing the driving springs.
Therefore, the stress on the interfaces $i > a$ relaxes, making a macroscopic multi-slip event unlikely.
A sequence of single-slip events is thus to be expected.
With an increasing number of layers, the slip sequence of the multiple interfaces may lose its periodicity.

\section{Numerical simulations}

\subsection{Numerical model}

We implement the model system shown schematically in (\cref{fig:1a}) into numerical simulations.
The numerical model consists of $n + 1$ identical elastic layers separated by frictional interfaces.
Following~\cite{deGeus2019}, we idealize the frictional contact problem in order to focus on the disorder in the shear response along the frictional interface.
We assume that the interface is disordered but perfectly flat.
Furthermore, we assume that contacts detach when a critical shear stress is reached.
As such, the system is macroscopically at zero temperature (contacts do not detach spontaneously due to random fluctuations caused by an external temperature).
Moreover, fluctuations of normal stress -- e.g.\ due to an inhomogeneous load \cite{Ben-David2011}, inhomogeneous pore pressures \cite{Harris1993}, or partial slip \cite{Ke2018} -- do not affect the contact strength.
In particular, we consider a mesoscopic scale on which an effective `block' of a finite width resists elastically to shear up to a threshold, after which it yields.
The local slip then propagates until a new `contact' is formed (i.e., it is again elastic but with a new threshold).
In this framework, each block represents a frictional contact (or a patch of contacts that are so strongly coupled by elasticity that they act as an effective contact) that, upon yielding, forms a new contact with a new yielding threshold.

Geometrically we do not seek to precisely model \cref{fig:1a} as its numerical treatment, together with the disorder, requires an intractably large number of blocks.
Therefore, we consider periodic boundary conditions in the horizontal direction and control the mean position of each slab through a parabolic potential energy.

The details of the numerical model are as follows: each frictional interface consists of $n_x$ equal-sized square blocks of linear size $l_0$ that are completely linear elastic under volumetric deformation but yield under shear (deviatoric deformation) when a set yield stress is reached.
Assuming that the yield threshold is isotropic in principal deviatoric strain space, this model now corresponds to a deviatoric potential energy that consists of a sequence of parabolic potentials in equivalent deviatoric strain space.
The disorder arises from independently randomly drawing the yield strain sequence of each block.
We assume that the blocks and the bulk have the same elastic moduli.

Differently from~\cite{deGeus2019}, we add a parabolic potential (with curvature $K$) to the mean horizontal position of each of the elastic slabs $i > 0$, adding a homogeneous force density per layer.
The bottom layer is not driven through its mean position; instead, the position of the bottom edge is fixed.
We set the mean horizontal position of a slab $i$ equal to $\gamma i h$, such that $\gamma$ represents the lever rotation (see \cref{fig:1a}).

A key feature of the model is that shear can be applied according to the quasistatic protocol.
Moreover, the linear elastic response permits an event-driven protocol.
In alternation, we increase the shear by a finite amount $\Delta \gamma(t)$ that is maximized with the constraint that no microscopic yielding takes place (preserving mechanical equilibrium), and then add an infinitesimal shear $\delta \gamma$ to trigger a microscopic event (after which we minimize energy).
This allows us to perfectly separate events.

We choose $n_x = 2 \times 3^6$, which is still tractable to simulate, but of the minimal order not to be dominated by finite size effects, as we checked for a single frictional interface~\cite{deGeus2019,deGeus2022}.
Furthermore, we take $h / \ell_0 \approx n_x / 4$ based on balancing $h / (\ell_0 n_x)$ small enough to have acoustic interactions while avoiding driving the blocks in a fixed displacement such that collective effects are suppressed if $h \ll n_x \ell_0$ (e.g.~\cite{Rosso2009}).

The above model predicts stick-slip behavior~\cite{deGeus2019} when full inertial dynamics are considered (using overdamped dynamics, this model predicts the abundantly studied depinning transition~\cite{Fisher1998}).
We consider such inertial dynamics by applying the finite element method to discretize space.
Along the frictional interface(s), elements coincide with the mesoscopic blocks.
In the elastic slabs away from the frictional interface, the elements are coarsened to gain numerical efficiency (such that the height $h$ is only approximated as we fix the aspect ratio of elements to one, see \cref{sec:SI:numerical}).
We use the velocity Verlet algorithm to integrate discrete time (with a time step significantly smaller than the time of a single oscillation in a well in one block).
We remark that assuming periodicity requires us to add a small damping term to the inertial dynamics such that waves with a wavelength equal to the horizontal size of the system ($n_x l_0$) are critically damped.
Consequently, we must take $h / \ell_0 < n_x$ to have acoustic coupling between the interfaces.

We note that following the underdamped dynamics in such a disordered system is very costly -- it requires on the order of one billion time steps per realization.
We are able to perform such simulations by using a dedicated optimized code, but mostly because we make two important assumptions.
First, we coarsen elements in the elastic bulk as we discuss above.
This is known to lead to spurious wave reflections \cite{Bazant1978}, but we presume that these are not important because disorder already leads to a broad spectrum of wavelengths.
Second, we assume small strains and rotations, such that the integration volume is assumed constant -- reducing the computational cost of numerical integration.
To still solve significant slips, the yield strains of each block are scaled by a small factor (all presented results have been renormalized to eliminate this factor).
However, this still inherently limits the description to moderate slips.
As such, we cannot choose an arbitrarily small driving stiffness $K$.

\subsection{Numerical results}

\begin{figure}[htp]
    \centering
    \subfloat{\label{fig:2a}}
    \subfloat{\label{fig:2b}}
    \subfloat{\label{fig:2c}}
    \subfloat{\label{fig:2f}}
    \includegraphics[width=\columnwidth]{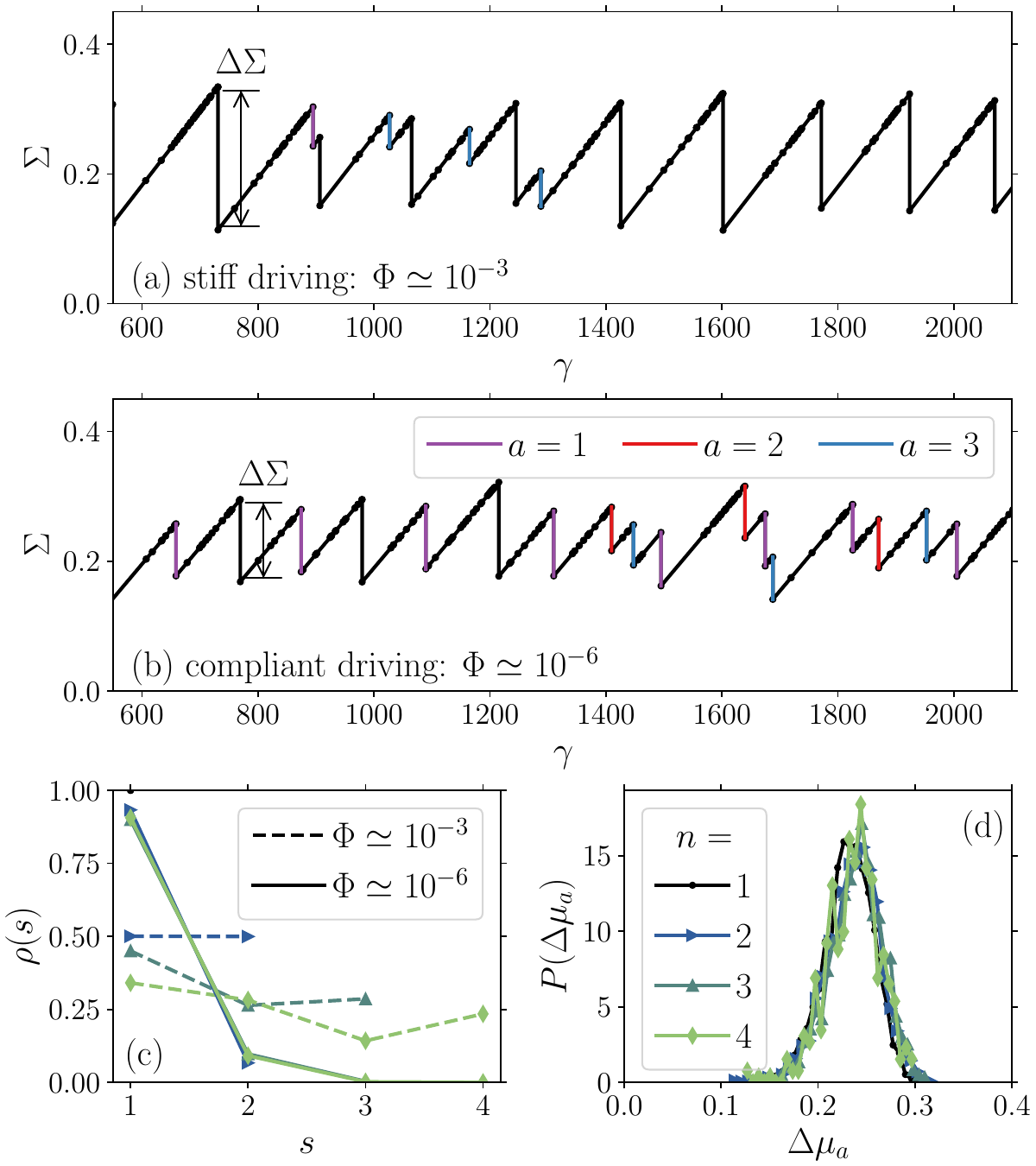}
    \caption{
        Numerical results: (a, b) Typical steady-state global stress $\Sigma$ response as a function of applied shear $\gamma$ for $n = 3$ for \protect\subref{fig:2a} stiff driving and \protect\subref{fig:2b} compliant driving.
        We indicate all (microscopic) yielding events with a black dot marker.
        Slip events on a single layer are indicated in color (see legend), while slip events in black involve more than one interface.
        \protect\subref{fig:2c}
        The fraction $\rho(s)$ of macroscopic slip events involving $s = 1, \ldots, n$ layers, for stiff (dashed) and compliant (solid) driving; see legend in \protect\subref{fig:2f} for color-map and markers.
        \protect\subref{fig:2f}
        Distribution of stress drops at the slipping interface for different $n$ in the compliant regime (for slip events on a single layer, for which $s = 1$ in \protect\subref{fig:2c}).
        See the main text for definitions and units.
    }
    \label{fig:2}
\end{figure}

Our numerical model allows us to first illustrate the simple argumentation on the role of driving that we made above.
We consider a driving rigidity such that $\Phi \simeq 10^{-3}$ (stiff driving) and $\Phi \simeq 10^{-6}$ (compliant driving).
In the two-dimensional model, $A = n_x \ell_0$, such that $K_s = 4G$ for our geometry; we use $K = 10^{-3}$ and $K = 10^{-6}$ and $G = 1 / 2$.
In \cref{fig:2a} and \cref{fig:2b}, for stiff and compliant driving, respectively, we plot a typical macroscopic stress $\Sigma$ (volume-averaged stress) as a function of applied shear $\gamma$.
Note that the stress is shown in units of the typical yield stress of one block, and the rotation in units of the rotation needed to yield a typical block at $i = 1$.

Macroscopic slip events are defined when all blocks along one or more layers yield at least once.
Below, we will refer to sliding interfaces and associated quantities by an index $a$, while $i$ will be kept as the running index for the layers.
Slip events correspond to macroscopic stress drops in \cref{fig:2a,fig:2b}, and we distinguish between `single-slip' events (all blocks on a single layer yield at least once) and `multi-slip' events (all blocks on more than one layer yield at least once).
Stress drops produced by single-slip events are labeled following the color code introduced in \cref{fig:1}, while multi-slip events are kept black.
These slip events are separated by `stick' intervals during which only microscopic events are observed, where one or several blocks yield at least once, as indicated with markers (black dots).

The results confirm that stiff driving causes a periodic stick-slip sequence with many slip events corresponding to multi-slip events (\cref{fig:2a}) while compliant driving results in a seemingly less periodic sequence of single-slip events in \cref{fig:2b}.
This finding is supported by plotting the fraction of slip events involving $s = 1, \ldots, n$ interfaces in \cref{fig:2c} for different $n$ (see legend in \cref{fig:2f}).
On the one hand, stiff driving results in single- and multi-slip events for a comparable fraction of loading history (we discuss in \cref{sec:SI:sequences} that sequences of single and multi-slip events alternate).
On the other hand, compliant driving shows single slip in the large majority of slip events.

In the compliant regime, a direct measurement of the stress drop along the slipping interface $a$, $\Delta \mu_a$, displays no $n$ dependence in \cref{fig:2f}.
The quantity $\mu_a$ is defined as the volume average stress on the blocks corresponding to weak layer $a$, also shown in units of the typical yield stress of one block.
Given that, by construction, normal stress plays no role in our model, here $\mu_a$ is akin to a friction coefficient.
The finite width of the distribution is attributed to the inherent disorder of the interfaces.

As we have shown that, for high-$\Phi$ (stiff driving), multilayer stick-slip is apparently similar to that of a single interface, next, we concentrate on the low-$\Phi$ regime to explore a potential influence of the number of plates $n$.

\section{Experiments}

We proceed by proposing an experimental realization of the sheared-multilayer model system of \cref{fig:1a}, adapted to measure the effect of the number of sliding layers $n$ on the slip synchronicity and amplitude; see \cref{fig:1b}.
Similarly to the numerical model, the position of each slab is driven by connecting it to the driving lever through linear springs (see schematic in \cref{fig:1a}).
Naturally, connecting the spring to the edges of the slabs might introduce boundary effects, but this is mitigated by the fact that our experimental system is effectively much larger than our numerical model (given that it presumably has much more local contact patches).

\subsection{Experimental apparatus}

\begin{figure}[h]
    \renewcommand\thesubfigure{(a.\arabic{subfigure})}
    \subfloat{\label{fig:setup:a:1}}
    \subfloat{\label{fig:setup:a:2}}
    \setcounter{subfigure}{0}
    \renewcommand\thesubfigure{(b.\arabic{subfigure})}
    \subfloat{\label{fig:setup:b:1}}
    \subfloat{\label{fig:setup:b:2}}
    \subfloat{\label{fig:setup:b:3}}
    \setcounter{subfigure}{0}
    \renewcommand\thesubfigure{(c.\arabic{subfigure})}
    \subfloat{\label{fig:setup:c:1}}
    \subfloat{\label{fig:setup:c:2}}
    \setcounter{subfigure}{0}
    \renewcommand\thesubfigure{(d.\arabic{subfigure})}
    \subfloat{\label{fig:setup:d:1}}
    \subfloat{\label{fig:setup:d:2}}
    \setcounter{subfigure}{0}
    \renewcommand\thesubfigure{(e.\arabic{subfigure})}
    \subfloat{\label{fig:setup:e:1}}
    \subfloat{\label{fig:setup:e:2}}
    \setcounter{subfigure}{0}
    \renewcommand\thesubfigure{(\alph{subfigure})}
    \subfloat{\label{fig:setup:a}}
    \subfloat{\label{fig:setup:b}}
    \subfloat{\label{fig:setup:c}}
    \subfloat{\label{fig:setup:d}}
    \subfloat{\label{fig:setup:e}}
    \centering
    \includegraphics[width=\columnwidth]{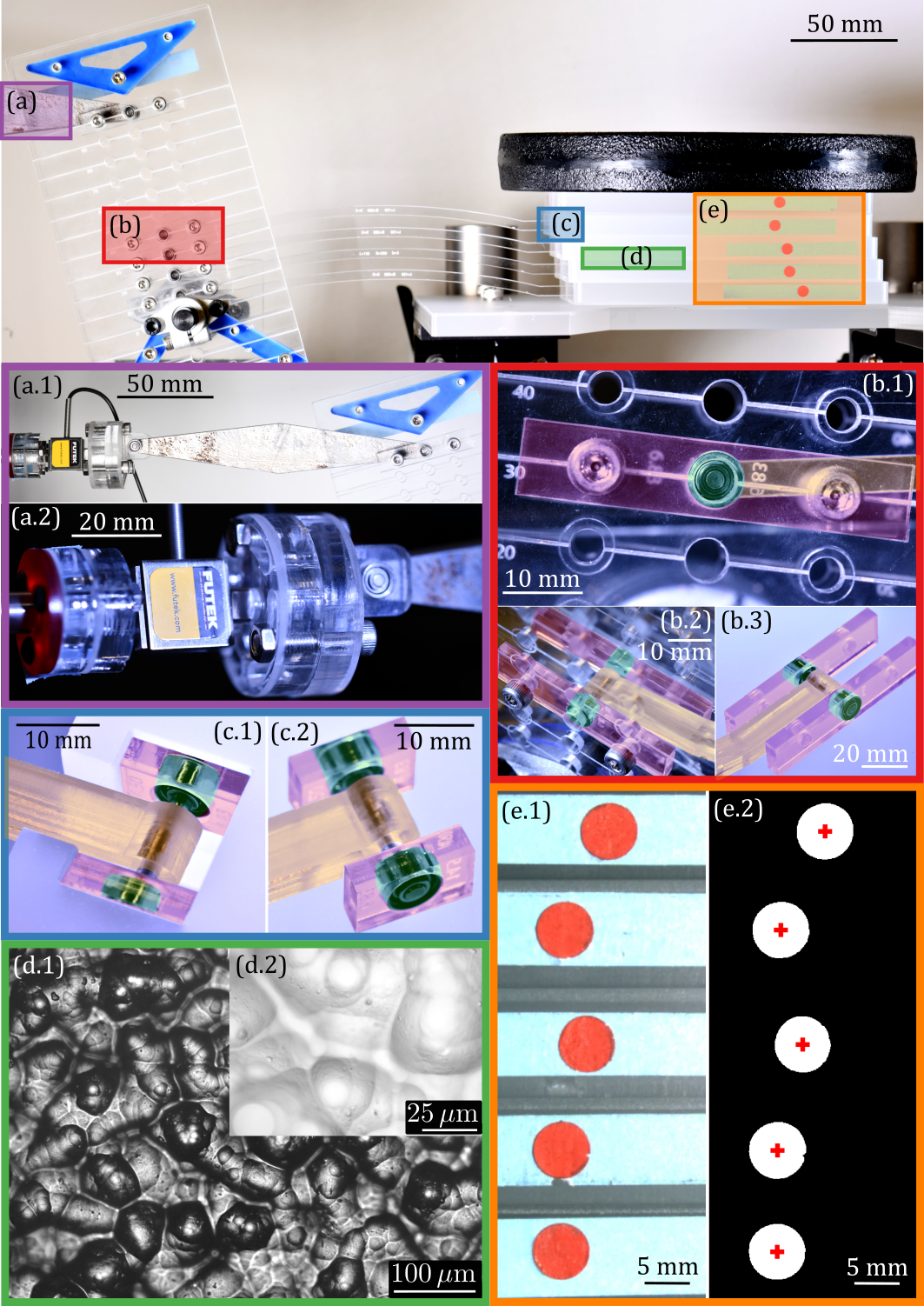}
    \caption{
        Technical details on the main components of the experimental apparatus.
        Top: overview of the set-up with the location of the different parts: (a) force measurement, (b) lever-spring link, (c) spring-slab link, (d) frictional slabs surface and (e) position measurement.
        Relevant details are shown in the bottom panels (numbered (a.1), etc.), see text for details.
    }
    \label{fig:setup}
\end{figure}

The experimental setup shown in \cref{fig:1b} comprises a stack of frictional plates (color-coded from purple to orange), an actuating lever (green), and driving springs (pink).
Each component of the setup is detailed in \cref{fig:setup} and below.

The stack is made of a set of rectangular PMMA slabs (Snow WH10 DC by R\"ohm), each of dimensions $h = 10\,\mathrm{mm}$, $L = 150\,\mathrm{mm}$ and an out-of-plane width of $80\,\mathrm{mm}$.
A normal force $N$ is applied on the topmost slab by a dead weight of $5\,\mathrm{kg}$ ($N = 49\,\mathrm{N}$).
To ensure a spatially homogeneous contacting surface at this relatively low normal force (compared to other PMMA-PMMA friction experiments~\cite{Ben-David2010,Svetlizky2014,Gvirtzman2021}), we use acrylic plates whose surface was pre-roughened with asperities of size $\sim25\,\mu\mathrm{m}$ that are larger than potential natural height variations of PMMA (\cref{fig:setup:d}).
We assume that the normal force is uniformly distributed and that it is the same for each layer (the weight of each slab is less than $3\%$ of that of the dead weight).

The stack is sheared by imposing the displacement at the top of the lever ($H = 100\,\mathrm{mm}$) at a constant speed $V = 10\,\mu\mathrm{m / s}$ (i.e.\ $\dot \gamma = V / H = 10^{-4}\,\mathrm{s}^{-1}$), using a DC linear actuator (L-220.70DG, Physiks Instruments) that is attached via a steel junction assumed rigid (\cref{fig:setup:a:1}).
The PMMA lever, made of two $6\,\mathrm{mm}$ thick and $100\,\mathrm{mm}$ wide parallel slabs, is sufficiently wide not to bend while pulling the slabs and rotates smoothly on ball bearings around its rotation axis.
The total horizontal force $F$ needed to rotate the lever (\cref{fig:1}) is measured using a uniaxial force sensor (LRM200 $25\,\mathrm{lb}$, Futek) placed between the steel junction and the actuator (\cref{fig:setup:a:2}).

The springs connecting the slabs to the lever are curved beams laser-cut from PMMA (colored in pink in \cref{fig:1b}), with an equivalent stiffness of $K = 55\,\mathrm{N / mm}$ when pulled or compressed along the horizontal axis.
The beams have a rectangular $5\times5\,\mathrm{mm}$ cross-section and are pre-curved with a radius of curvature of $100\,\mathrm{mm}$ over a cord of $150\,\mathrm{mm}$ in the middle and extended on both sides by two $25\,\mathrm{mm}$ straight portions.

The ends of the springs are attached to both the slabs and lever via ball-bearing links to ensure a free rotation and, thus, horizontal driving forces.
The links are inserted between the two slabs of the lever and are made of a central steel pin press-fitted at the end of the spring (\cref{fig:setup:b:1}, orange).
This pin is embedded into small ball bearings on each side (green), allowing a free relative rotation between the springs and the lever.
The bearings are then press-fitted into acrylic components (pink) with two threaded holes.
These components are screwed directly on both sides of the lever.
The rotation axis of the pins is aligned precisely along a line passing through the center of rotation of the lever.
\cref{fig:setup:b:1,fig:setup:b:2} show the side and top-side views of the link; and \cref{fig:setup:b:3} shows the link prior to attachment to the lever.
The other sides of the springs are linked to the frictional slabs.
The spring-slab links have a similar design, but here the acrylic components are press-fit into the slabs.
The height of the acrylic components is slightly smaller than the height of the slabs ($8\,\mathrm{mm}$ vs.~$10\,\mathrm{mm}$) to avoid any contact with the interfaces.
\cref{fig:setup:c:1} and \cref{fig:setup:c:2} show top-side views of the links, respectively, attached and out of a frictional slab.
These links ensure a rigid connection between all the components, so that only the springs deform during actuation.
Moreover, the ball bearings ensure that the force transmission remains horizontal as the lever rotates.
The steel junction between the lever and force sensor is also attached via analogous links to remain horizontal at all times.

The experimental set-up, with its springs and PMMA slabs, corresponds to the compliant driving limit with $\Phi \simeq 6 \times 10^{-5}$, of the same order as for the compliant regime in the numerics.
Indeed, $\Phi \equiv K / K_s$, with $K_s = A G / h$ the shear stiffness of the slabs and $G \equiv E / (2 (1 + \nu))$, with, for PMMA, Young's modulus $E = 2 \, \mathrm{GPa}$ and Poisson's ratio $\nu = 0.3$.

In addition to the global force measurement $F$, we also measure the absolute average horizontal position of each slab $x_i$, by tracking a red marker placed on their side (\cref{fig:setup:e:1}), from photographs taken at a rate of $5\,\mathrm{fps}$ using a digital camera (Flea3 FL3-U3-20E4C, Flir, linear pixel size: $70\,\mu\mathrm{m}$).
Using a color threshold, the red markers are extracted, as shown in \cref{fig:setup:e:2}, and their geometric center is located (red crosses).
Owing to the large size of the markers ($154\,\mathrm{mm}^2$, corresponding to approximately $30 000$ pixels), $x_i$ can be determined with a sub-pixel resolution of $5\,\mu\mathrm{m}$.
The relative displacement between slabs is $R_i \equiv x_i - x_{i - 1}$ (see \cref{fig:1a}), which serves as a proxy for the total slip at the interface $i$ (neglecting the shear deformation of the slabs).

To vary the number of sliding interfaces $n$, we keep the same number of slabs ($5$) but remove $5 - n$ springs, starting from the top (see \cref{fig:1b} where $n = 4$).
This procedure ensures robust image detection and reduces external contamination of the interfaces by keeping them in contact.
Each time the slabs are disassembled to vary $n$, the interfaces are cleaned with isopropanol and quickly dried using compressed air.
For each value of $n$, we perform $10$ runs during which we drive over a range $\Delta\gamma = 0.6\,\mathrm{rad}$, starting at $\gamma = -0.30\,\mathrm{rad}$, each time excluding $\gamma$ between $-0.3\,\mathrm{rad}$ and $-0.27\,\mathrm{rad}$ ($300\,\mathrm{s}$) to ensure measuring in a steady state.
After each run, the lever is reset back to $\gamma = -0.30\,\mathrm{rad}$.
On average, each connected layer is forced to move by a total relative distance of $R_\mathrm{tot} = h \Delta\gamma = 6\,\mathrm{mm}$ during a run that lasts $6000\,\mathrm{s}$ in total.

\subsection{Measurements and lever kinematic}

\begin{figure}[htp]
    \subfloat{\label{fig:3a}}
    \subfloat{\label{fig:3b}}
    \subfloat{\label{fig:3c}}
    \centering
    \includegraphics[width=\columnwidth]{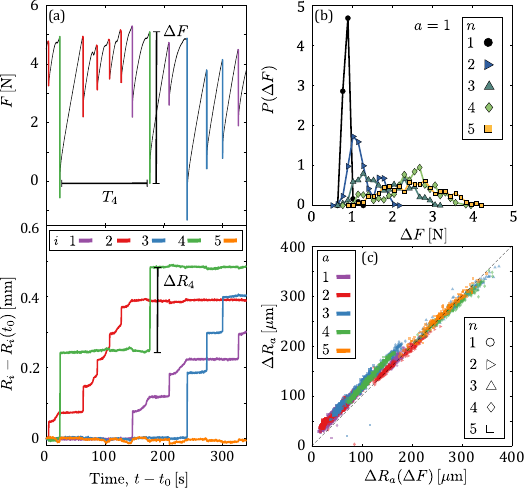}
    \caption{
        (a) Top plot: extract of a time series of the macroscopic force $F(t)$ for a system of $n = 4$ frictional interfaces.
        The color of the force drops $\Delta F$ follows the color code in \cref{fig:1} and indicates the index $a$ of the slipping interface.
        Bottom plot: corresponding relative displacement (total slip) $R_i(t) - R_i(t_0)$ (with $t_0 = 5400\,\mathrm{s}$ an arbitrary value) of each interface $i$.
        Each slip event is characterized by $\Delta R_a$.
        We denote $T_a$, the time between subsequent slip events on the same interface.
        Note that we show $i = 5$ only for completeness, by definition, $R_5 = 0$ if $n = 4$.
        (b) Probability distribution function $P(\Delta F)$ for slip at interface $a = 1$ for an increasing number of layers~$n$.
        (c) Comparison between a direct measurement of $\Delta R_a$, and the computed $\Delta R_a(\Delta F)$, obtained through \cref{eq:DR_DF_multi}, for each detected slip event.
    }
    \label{fig:3}
\end{figure}

For a stack with $n = 4$, we present in \cref{fig:3a} a typical time series extract of the force $F(t)$ required to actuate the lever (top left plot), together with the corresponding relative position of the slabs $R_i$(t) (bottom-left plot).
The experiments exhibit stick-slip, with stick periods when the slabs are immobile ($R_i \approx \mathrm{constant}$) and $F$ increases monotonically, punctuated by macroscopic slip events.
These slip events are identified by a sudden position jump, $\Delta R_a > 0$ (with $a$ denoting the sliding interface), accompanied by an abrupt force drop $\Delta F > 0$, cf.~\cref{fig:3a}.
On all occasions, we find that only one layer slips at a time, recovering similar dynamics as in the numerical model in the compliant driving regime with a similar value for $\Phi$ (\cref{fig:2b}).
However, we note that during the stick periods, we observe what seems to be `slow slip' where an interface moves gradually, leading to a non-linear force response.
These are out of our primary focus but are discussed at the end of the section.

For each value of $n$, we acquire an ensemble of at least $100$ slip events per layer, such that the slip quantities associated can be represented as probability distributions.
For example, in \cref{fig:3b}, we show the probability distribution of force drops, $P(\Delta F)$, occurring on the interface $a = 1$, for all cases of $n$ considered.
Starting from the peaked distribution for $n = 1$, as $n$ increases, the distributions broaden and take higher average values.

In contrast with more classic stick-slip experiments with a single interface~\cite{Heslot1994}, the global measure $\Delta F$ is not a direct quantification of the frictional properties of the interface but couples with the specific kinematic of the lever.
Still, the fact that only one interface slips allows us to extract a jump in a friction-like quantity $\Delta \mu_a$ (or stick-slip amplitude) from $\Delta F$.
We define the friction coefficient $\mu_i$ of an interface $i$ as the horizontal force acting on this interface divided by the normal force.
Considering the horizontal force balance on an interface $i$, the interface has to resist the combined forces of the pulling springs of the slabs $j \geq i$, such that
\begin{equation}
    \label{eq:mu_a_def}
    \mu_i = \sum_{j = i}^n \frac{f_j}{N},
\end{equation}
where $f_j$ is the force due to the driving spring on slab $j$ (see \cref{fig:1a} for a visual representation of $\mu_i$ and $f_j$).
When an interface $a$ slides by $\Delta R_a$, the relative positions of the other interfaces remain unchanged (we only observe single-slip events).
Consequently, the absolute horizontal position $x_i$ of the layers above $a$ is increased by $\Delta R_a$:
\begin{equation}
    \label{eq:slip}
    \Delta x_i =
    \begin{cases}
        \Delta R_a > 0 \quad & \text{if} \quad i \geq a \\
        0 \quad & \text{if} \quad i < a
    \end{cases}
    .
\end{equation}
This sliding induces a drop in the spring forces: $\Delta f_i = K \Delta x_i = K\Delta R_a$ for $i \geq a$, and $\Delta f_i = 0$ for $i < a$.
Note that for consistency, we define $\Delta f_i$ to be positive.
From \cref{eq:mu_a_def}, we can then express $\Delta\mu_a$ as a function of the slip distance $\Delta R_a$:
\begin{equation}
    \label{eq:mu_a_DR}
    \Delta\mu_a = \frac{K}{N}(n - a + 1)\Delta R_a.
\end{equation}

We proceed by linking $\Delta R_a$ to the global force drop $\Delta F$, using the fact that only one interface slips at a time.
Through moment balance on the lever, we obtain:
\begin{equation}
    \label{eq:moment}
    F = \sum_{i = 1}^n f_i \frac{ih}{H}.
\end{equation}
Combining \cref{eq:slip,eq:moment}, we obtain a relation between the global quantity $\Delta F$ and the local $\Delta R_a$:
\begin{equation}
    \label{eq:DR_DF_multi}
    \Delta F = \sum\limits_{i = a}^n K\Delta R_a\frac{ih}{H} = K\Delta R_a\frac{h}{H}\frac{(n + a)(n - a + 1)}{2} .
\end{equation}
(Note that $i$ is the only varying term in the sum and $\sum_{i = a}^n i = (n(n + 1) - a(a + 1)) / 2 = (n + a)(n - a + 1) / 2$.) This result is verified in \cref{fig:3c}.
Indeed, a direct measurement of $\Delta R_a$ is very close to the inversion of \cref{eq:DR_DF_multi}, in which $\Delta R_a$ follows from the measured $\Delta F$ without any fitting parameter.

Finally, we combine \cref{eq:mu_a_DR,eq:DR_DF_multi} to obtain the sought relation between $\Delta \mu_a$ and $\Delta F$:
\begin{equation}
    \label{eq:mu_a}
    \Delta\mu_a = \frac{H}{h}\frac{2}{(n + a)}\frac{U\Delta F}{N}.
\end{equation}
We have thereby disentangled the friction properties of the interface from the kinematics of the lever.
Using \cref{eq:mu_a}, we can now obtain a measure of the stick-slip amplitude of the interfaces $\Delta \mu_a$, extracted directly from the global force $\Delta F$.
The position measurements $R_i$ are used only to identify the slipping interface $a$.

The lever kinematics introduces a strong coupling between the interfaces.
Via \cref{eq:mu_a_def,eq:slip}, a slip on an interface $a$ will induce drops of friction coefficient $\mu_i$ on all the other interfaces, even if no slip occurs on them.

\begin{figure}[htp]
    \subfloat{\label{fig:4a}}
    \subfloat{\label{fig:4b}}
    \subfloat{\label{fig:4c}}
    \subfloat{\label{fig:4d}}
    \centering
    \includegraphics[width=\columnwidth]{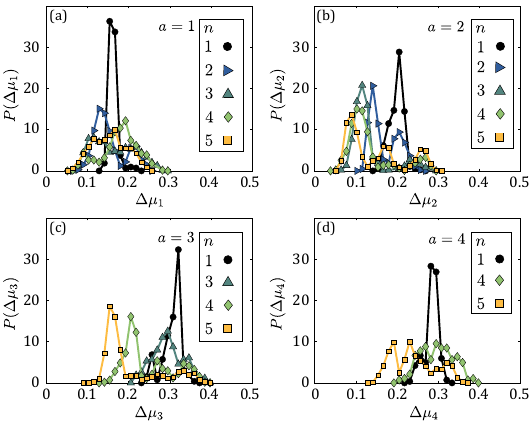}
    \caption{
        Probability distribution functions of the stick-slip amplitude $P(\Delta \mu_a)$ as a function of $n$ for the different interfaces: (a) $a = 1$, (b) $a = 2$, (c) $a = 3$ and (d) $a = 4$.
    }
    \label{fig:4}
\end{figure}

\textit{Central experimental result.}
Next, we assess the effect of having multiple sheared interfaces on their frictional properties.
\cref{fig:4} shows the probability distributions $P(\Delta \mu_a)$ associated with the different sliding interfaces $a$ (different panels) and the increasing number of total active interfaces $n$ (different colors).
Each interface is compared to its response when sliding individually ($n = 1$ in black, see \cref{sec:SI:individual} for experimental protocol).
For all the interfaces, the stacks exhibit significantly enriched statistics when compared to a single sliding layer ($n = 1$), in contrast to the numerical predictions reported above (\cref{fig:2c}) where $\Delta\mu_a$ was independent of $n$.
With increasing $n$, the location of the major peak of $P(\Delta \mu_a)$ shifts to lower values of $\Delta \mu_a$ and the respective distributions become broader as secondary peaks emerge.

\subsection{Interpretation}

We seek to interpret the above experimental findings evidencing a variation of the frictional properties as more layers are added to the stack (\cref{fig:4}), whereas they are independent of $n$ in the numerics (\cref{fig:2f}).
Our methodology is to identify the possible mechanisms contributing to variations of $\Delta \mu_a$ and quantify their potential signature in the physical and numerical experiments.
First, we will attribute the finite width of the peaks in the $P(\Delta \mu_a)$ distributions, even for $n = 1$, to the interface disorder also present in the numerics.
Then, we will argue that the major changes with $n$ observed in $P(\Delta \mu_a)$ result from the combined effect of increasingly complex loading paths and aging of the interfaces.
Finally, we speculate how creep of the interfaces, at the origin of aging and slow slip, could be influenced by the increase of an effective temperature with $n$.

\textit{Interface disorder.}
Even when sliding individually ($n = 1$), the frictional properties of the interfaces are distributed: $P(\Delta \mu_a)$ has a finite width, see black curves with circles in \cref{fig:4}.
These underlying statistical fluctuations, also present in the numerical model (\cref{fig:2f}), are considered to be related to the disorder of the contacting interfaces.
The rough interface induces a broad distribution of barriers, leading to collective events with non-trivially distributed sizes.
These collective events nucleate the macroscopic slip~\cite{deGeus2019,deGeus2022}, such that the stress at which slip is nucleated is distributed.

\begin{figure}[htp]
    \subfloat{\label{fig:5a}}
    \subfloat{\label{fig:5b}}
    \subfloat{\label{fig:5c}}
    \subfloat{\label{fig:5d}}
    \centering
    \includegraphics[width=\columnwidth]{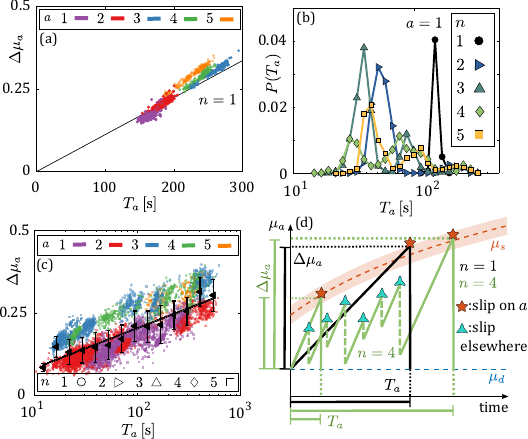}
    \caption{
        \protect\subref{fig:5a}
        For an interface sliding individually ($n = 1$), stick-slip amplitude $\Delta \mu_a$ as a function of the waiting time since the last slip event $T_a$.
        The black line corresponds to the prediction of \cref{eq:disorder}.
        For multiple sliding interfaces ($n \geq 1$):
        \protect\subref{fig:5b}
        Probability distribution of the waiting time $T_a$ between two consecutive slip events at interface $a = 1$, for increasing $n$.
        \protect\subref{fig:5c}
        For each detected slip event, correlation between $\Delta \mu_a$, and its corresponding waiting time $T_a$ (semi-logarithmic scale).
        The black markers correspond to the mean values for a logarithmic binning of $T_a$ (error bars indicate the standard deviation for that bin), and the dotted line a fit of \cref{eq:aging} with $B = 0.053 \pm 0.005$.
        Alternative representations of this data-set are displayed in \cref{sec:SI:aging_variations}.
        \protect\subref{fig:5d}
        Schematic of the proposed mechanism leading to multimodal and wider distributions of $T_a$ (and thus $\Delta \mu_a$) as $n$ increases.
    }
    \label{fig:5}
\end{figure}

Let us now verify experimentally that, in the case of individually sliding layers ($n = 1$), the measured fluctuations of $\Delta \mu_a$ correspond to distinct frictional interface strength.
In the individual configuration, the spring drives the layer at a constant rate $\dot f_a = K h \dot \gamma$ (in practice, the shear rate imposed by the lever is adapted for each $a$ and set to $\dot{\gamma} / a$ to account for the difference in height, see \cref{sec:SI:individual}).
The shear applied to the interface then grows at a rate $\dot \mu_a = \dot f_a / N = K h \dot \gamma / N$.
As such, we expect the stick-slip amplitude to be proportional to the time between consecutive slips $T_a$, following:
\begin{equation}
    \Delta \mu_a = T_a K h \dot \gamma / N.
    \label{eq:disorder}
\end{equation}
This is consistent with our data in \cref{fig:5a}, thus confirming that the finite width of the primary and secondary peaks in $P(\Delta \mu_a)$ results from statistical fluctuations of frictional interface strength.
However, these fluctuations do not account for the shifts of the peaks and the appearance of secondary peaks.

In the following, we argue that coupling via the lever is responsible for increasingly complex loading paths with $n$, leading to broadly distributed waiting times between slips $T_a$ (I).
Then, aging of the interface, present in the experiments but not in the numerics, becomes significant in broadening $\Delta \mu_a$ (II).

\textit{(I) Complex loading path.}
The coupling of the interfaces via the driving lever has two effects.

First, the loading rate at a given interface increases with $n$.
Using \cref{eq:mu_a_def} and $\dot f_i = K i h \dot\gamma$ while no interfaces are sliding, an interface $a$ in a stack of $n$ layers undergoes a loading rate of $\dot \mu_a = K \dot \gamma h (n + a)(n - a + 1) / (2 N)$, which is an increasing function of $n$.
With this increased loading rate with $n$, we thus expect the time between slips $T_a$ to typically decrease with $n$ for all interfaces (note that this effect is not captured by our quasistatic numerics in which $\dot{\gamma} = 0$).

Second, the lever couples the load on an interface to slips on other interfaces via \cref{eq:slip,eq:mu_a_def}.
Slip on one interface unloads the interface itself, but also all the other interfaces in the stack, even if no slip occurs on them.
In particular, if no slip occurs anywhere, $\mu_a$ is a linear function of time and reaches its frictional strength with a time $T_a$, which is a decreasing function with $n$.
But as $n$ increases, slips occurring on other interfaces decrease $\mu_a$ instantaneously (with a $K$- and $n$-dependent amplitude \footnote{In our numerics, there is no n-dependence due to the use of a force \emph{density}.}).
This delays the time when the interface reaches its frictional strength, leading to various $T_a$.

In \cref{fig:5b}, we plot the probability distribution function of $T_a$ for $a = 1$ and increasing $n$.
We indeed recover that the peaked distribution for $n = 1$ shifts to lower values of $T_a$ with increasing $n$ (when loading the interface without interruption by other slips).
Moreover, secondary peaks in $P(T_a)$ start to appear, which we interpret to be due to slip events on the other layers.
These observations are robust for the other interfaces (see \cref{sec:SI:distributions}).
The change in loading rate with $n$, together with the complex loading path, allows our experimental system to probe a broad distribution of $T_a$ on all its interfaces.

\textit{(II) Aging.}
It is a known experimental fact that the macroscopic stress required for the onset of sliding, characterized by $\mu_s$ (the `static friction coefficient' in Amontons-Coulomb's terminology~\cite{Baumberger2006,Popova2015}), depends on the duration $T$ that the interface was static: $\mu_s = B \ln (T / T_0)$~\cite{Heslot1994,Baumberger2006,Dokos1946,Dieterich1972}, where the aging rate of the interface $B$ remains a constitutive parameter and $T_0$ a microscopic time-scale.
Let us now consider $\Delta \mu_a$ as a proxy for $\mu_s$, assuming that a slip event unloads the interface to a well-defined and constant quantity ($\mu_d$ the `dynamic friction coefficient'), as is supported by~\cite{Heslot1994}.
We expect to find, for each sliding interface $a$ and over a wide range of $T_a$, that the stick-slip amplitude follows:
\begin{equation}
    \Delta \mu_a = B \ln (T_a / T_0).
    \label{eq:aging}
\end{equation}
Thereby $T_0$ is an unknown microscopic time-scale that we take equal to one second following common experimental literature (summarized e.g.\ in \cite{Baumberger2006}).
In \cref{fig:5c}, we assess this expectation experimentally by plotting $\Delta \mu_a$ versus $T_a$ in a semi-logarithmic scale.
To capture the general trend, we bin logarithmically $T_a$, corresponding to the black markers with corresponding error bars in \cref{fig:5c}.
These averaged values of $\Delta \mu_a$ are indeed consistent with a straight line in the semi-log plot, and we extract the slope $B = 0.053 \pm 0.005$, which is of the same order of magnitude as measured in classical stop-and-go experiments that are in the $10^{-2}$ order~\cite{Baumberger2006}, and of a direct surface observation on PMMA at room temperature that reports $B = 0.009 \pm 0.001$~\cite{Ben-David2010}.
Aging of the interface's contacts then translates the large peak shifts and the emergence of new ones in the $T_a$ distributions (\cref{fig:5b}) into qualitatively similar changes in the $\Delta\mu_a$ distributions as observed in \cref{fig:4}.

In \cref{fig:5d}, we schematically represent the coupled role of (I), (II), and disorder, following the evolution of the interfacial stress of an interface $\mu_a$ with time, starting from the last slip event.
The layer slips when it reaches $\mu_a = \mu_s (T_a)$, whereby $\mu_s$ is distributed in some way for fixed $T_a$ because of disorder (illustrated as a red-shaded area, where for simplicity, we lump all fluctuations in the threshold to sliding) and increases logarithmically with time because of aging.
For $n = 1$ (black line), $\mu_a$ increases linearly at the same rate for all the events, thus exploring a narrow region of $\mu_s$ (the shaded red region due to disorder) and $T_a$, following \cref{eq:disorder}.
In the case of multiple active interfaces ($n > 1$, green lines), $\mu_a$ increases faster given that $\dot \mu_a$ is an increasing function of $n$, and several scenarios arise.
If no other slip events occur in the stack, $\mu_a$ linearly reaches $\mu_s$, resulting in a lower value of $T_a$ and $\Delta \mu_a$.
However, if sliding events occur elsewhere during loading, $\mu_a$ will drop before linearly increasing again, delaying slip and thus increasing $T_a$, and consequently $\Delta\mu_a$, because of aging.

\begin{figure}[htp]
    \subfloat{\label{fig:6a}}
    \subfloat{\label{fig:6b}}
    \subfloat{\label{fig:6c}}
    \subfloat{\label{fig:6d}}
    \centering
    \includegraphics[width=\columnwidth]{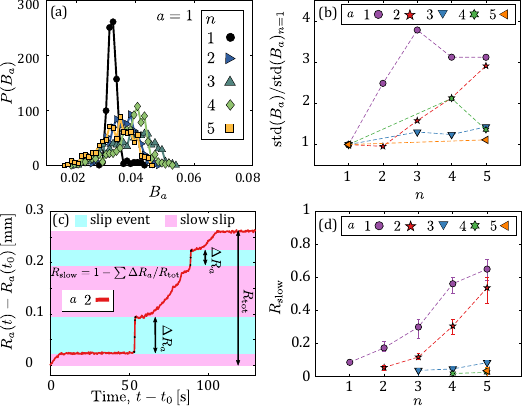}
    \caption{
        \protect\subref{fig:6a}
        Distribution of the aging rate $B_a = \Delta \mu_a / \ln (T_a / T_0)$ for all events on the interface $a = 1$ and all $n$.
        \protect\subref{fig:6b}
        Standard deviation of $B_a$ for all layers $a$, normalized by that quantity for $n = 1$, as a function of $n$.
        \protect\subref{fig:6c}
        For $n = 5$, a typical time evolution displaying slow slip of the relative position $R_a(t) - R_a(t_0)$, of the layer $a = 2$.
        The sliding distance $R_\mathrm{tot}$ is categorized as slip events of amplitude $\Delta R_a$, and slow slip.
        The fraction of slow slip is defined by $R_\mathrm{slow} = 1 - \sum \Delta R_a / R_\mathrm{tot}$.
        \protect\subref{fig:6d}
        Fraction of slow slip $R_\mathrm{slow}$ for all layers $a$ as a function of the number of layers $n$.
        Error bars represent the minimal and maximal values of $R_\mathrm{slow}$ obtained over $10$ experimental runs.
    }
    \label{fig:6}
\end{figure}

\textit{Creep effective temperature.}
During stick intervals, microscopic events occur on the interfaces, propagating elastic waves across the system \cite{deGeus2019}.
As we increase the number of interfaces in the system, we can expect that the overall mechanical noise created by the microscopic events also increases.
If we speculatively interpret this mechanical noise as an effective temperature, we would expect a change of aging rate $B$ with $n$.
Let us define the aging rate for a single event as $B_a \equiv \Delta \mu_a / \ln (T_a / T_0)$ and associated probability distributions for all associated events (see \cref{fig:6a} for $a = 1$, and \cref{sec:SI:distributions} for the other interfaces).
Although the mean of $P(B_a)$ does not change with $n$, we do find that the width of the distribution $P(B_a)$ is an increasing function of $n$ mainly for the lowermost interfaces ($i \leq 2$), as shown in \cref{fig:6b}.

For the same interfaces ($i \leq 2$), we also observe distinctly different slip dynamics when $n > 1$.
In particular, as $n$ increases, we find that interfaces $i = 1$ and $i = 2$ are increasingly subject to slow slip, defined as sliding significantly slower than the slip events (see \cref{fig:6c} for an example and \cref{sec:SI:slow} for a quantitative characterization).
Slow slip is not accompanied by a sudden macroscopic stress drop but rather by a slow decrease of $\dot{F}$.
In \cref{fig:6d}, we estimate the fraction of slow slip $R_\mathrm{slow}$ compared to the total sliding distance $R_\mathrm{tot}$.
It is computed by comparing $R_\mathrm{tot}$ to the total accumulated slip during individual slip events (sum of all slip events $\Delta R_a$), such that $R_\mathrm{slow} \equiv 1 - \sum \Delta R_a / R_\mathrm{tot}$.
Once a slow slip event starts, it appears to be stopped only by slip events occurring either on the same interface or on any other interface.
An increase in the effective temperature of the interface with $n$ could also act as a potential destabilization factor of the contacts at the interfaces, increasing the occurrence of slow slips with $n$.

\section{Discussion and Conclusion}

\subsection{Summary}

We have explored the stick-slip response of a system with multiple interfaces by proposing a model system comprising $n$ vertically stacked slabs, each connected to a lever whose rotation is imposed.
The interfaces were driven in quasistatic (homogeneous) shear.
We proposed a dimensionless quantity $\Phi$ as the ratio between the driving stiffness and the elastic shear stiffness of the slabs.
We have argued and demonstrated numerically that the system displays synchronization if $\Phi$ is sufficiently large ($\Phi \gtrsim 10^{-3}$).
In that case, the system acts close to a single frictional interface with effective properties.
If $\Phi$ is small ($\Phi \sim 10^{-6}$), interfaces slip one by one, as also confirmed experimentally.

We expect non-trivial collective effects with increasing $n$ only in the low-$\Phi$ limit, which we addressed through experiments.
In the numerics, the stick-slip amplitude of the interfaces $\Delta\mu_a$ displays a distribution with finite width because of statistical fluctuations of the interfaces, but no measurable changes with $n$.
By contrast, we measured experimentally that the probability distribution of stick-slip amplitude $\Delta\mu_a$ shows a general broadening with $n$, with peaks shifting to lower values and secondary peaks appearing.
The interfaces are coupled via the lever, exposing them to a complex loading path, and leading to a broad distribution of the waiting times $T_a$ between two slip events on an interface.
We find that $T_a$ is now spanning two decades, such that the aging of the interfaces plays a crucial role in the broadening of $\Delta\mu_a$.
The complex distributions of $\Delta\mu_a$ can then be interpreted as the combined effect of interface disorder, also observed numerically, and aging.
For narrow waiting times $T_a$, multiple slips explore the statistical fluctuations of the contacting interfaces, giving the finite width of the peaks in the distributions.
In addition, creep-induced aging gives a long-time general trend over widely distributed $T_a$.

Furthermore, we observe that increasing $n$ has a significant impact on creep-induced phenomena like aging rate and slow slip.
We suggest that these additional consequences of adding more layers to the stack might be evidence of an increase in an effective interface temperature due to the mechanical noise of microscopic events.

In conclusion, the relative rigidity of the drive against the layers dictates whether a stack of interfaces responds synchronously or not.
When layers slide one by one, increasing their number leads to a complex coupling, making the prediction of the next slip more challenging.

\subsection{Limitations and outlook}

It is pertinent to discuss some limitations of our model system and provide suggestions for future work.

\textit{Stiffness ratio.}
We have defined the stiff and compliant driving regimes, as characterized by the relative order-of-magnitude estimation of the respective stiffness ratio $\Phi$.
Identifying an equivalent of $\Phi$ in systems with more intricate geometries, such as fault networks, might contribute to clarifying their dynamics and help slip predictions.
Hence, a more systematic exploration of the response of stacks with varying $\Phi$ would be of great interest.

However, we are currently restricted to a limited range of $\Phi$.
For our numerics, low-$\Phi$ are challenging due to a combination of the assumption of small rotations and finite machine precision.
To be able to continue sliding indefinitely, our model should be extended with the possibility to reset the local deformation along the frictional interface to a zero average while keeping the identical stress state.
In contrast, high-$\Phi$ are challenging experimentally with the current setup.
If the driving stiffness is increased, the motor/lever system can no longer be considered rigid, invalidating the relation between global ($\Delta F$) and local ($\Delta\mu_a$) force drops (\cref{eq:DR_DF_multi}).
Instead, if the shear stiffness of the slab materials is significantly lowered, a different class of frictional properties is expected as soft materials tend to be adhesive~\cite{Viswanathan2017}.

\textit{Creep.}
Our proposed model system allowed us to measure the aging rate $B$ of the interfaces thanks to complex stick-slip sequences without the need for stop-and-go experiments.
However, while our measured value of $B$ is compatible with previous experiments on PMMA \cite{Baumberger2006,Ben-David2010}, it differs by a factor of five.
Possible sources of differences are roughness, inter-realization variations, and stress inhomogeneities.
First, the PMMA plates used here have a much higher surface roughness than in \cite{Baumberger2006,Ben-David2010}.
If our model is extended with thermal fluctuations it likely displays creep such that the relationship between the distribution of barriers, linked to surface roughness, and creep could be investigated.
Second, we measure $B$ on an ensemble of $n = 5$ interfaces.
Between interfaces it is estimated that $B$ differs by a factor of about two.
Third, recent experimental observations find a relationship between the aging rate $B$ and the applied shear load~\cite{Dillavou2020}.
Our setup naturally imposes a broad range of shear loads on the interface.
However, to measure the empirical law by \cite{Dillavou2020}, our setup would need to be augmented to also provide stress measurements per interface by measuring individually the internal forces $f_i$ of the driving springs.
To study this effect numerically, on top of adding temperature to capture creep, the model would likely have to be made sensitive to pressure inhomogeneities that may arise from the normal load or partial slip events, commonly thought to be a significant perturbation \cite{Ke2018}.

\textit{Slow slip.}
The origin of the experimentally observed slow slip is unknown.
A tempting hypothesis is that smooth sliding is the result of activated yielding events due to increasing mechanical noise on other interfaces.
However, it is not clear why this interpretation would lead to slow slip occurring predominantly on the lowermost interfaces (which is qualitatively robust to changing the order of the slabs, see \cref{sec:SI:slow}).
Furthermore, slow slip is not observed numerically.
This could, however, in part be due to our small homogeneous background damping term (currently chosen to avoid nonphysical periodic wave propagation).
An alternative hypothesis is that, by increasing $n$, the loading rate becomes sufficiently high to drive the interfaces away from the stick-slip regime (if $v \geq v_c$).
This second interpretation is consistent with slow slip being predominantly observed on the lowermost interfaces.
Note that, different from the experiments, our numerical model drives infinitely slowly.

\textit{Aftershocks.}
Aftershocks appear if creep is added to the drive of a simple spring-block model~\cite{Jagla2014,Houdoux2021}.
A creeping drive is often associated with the high temperatures in Earth's core~\cite{Jagla2014}.
Our experimental system displays slow sliding of `deep' layers already at room temperature.
A key question is if aftershocks appear in the top layers of our system as well.
Answering this question experimentally would require exposing microscopic events, which would likely involve studying acoustic emissions (for which PMMA may not be the optimal choice).

\textit{Acknowledgments.}
The authors thank Mathias Lebihain and Federica Paglialunga for fruitful discussions, and Lebo Molefe for providing the microscope images of the surface of the slabs.
S.P.~acknowledges financial support from the Japanese Society for the Promotion of Science as a JSPS International Research Fellow.
T.G.~acknowledges support from the Swiss National Science Foundation (SNSF) by the SNSF Ambizione Grant PZ00P2{\_}185843.

\bibliography{library}

\clearpage
\appendix

\section{Individual sliding}
\label{sec:SI:individual}

\begin{figure*}[htp]
    \centering
    \subfloat{\label{fig:SI:single:a}}
    \subfloat{\label{fig:SI:single:b}}
    \subfloat{\label{fig:SI:single:c}}
    \includegraphics[width=0.8\textwidth]{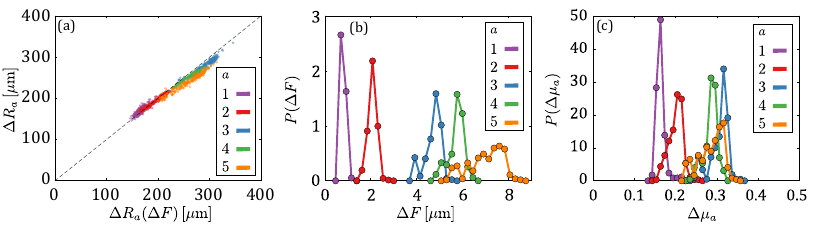}
    \caption{
        \protect\subref{fig:SI:single:a}
        Verification of $\Delta R_a(\Delta F)$ (\cref{eq:SI:DR_DF_single}) for individual interfaces.
        \protect\subref{fig:SI:single:b}
        The probability distributions of the drop of force $\Delta F$ for the different interfaces $a$ are shown using a different color and marker.
        \protect\subref{fig:SI:single:c}
        Probability distributions of the corresponding drop in frictional resistance $\Delta \mu_a$ as estimated from $\Delta F$ using \cref{eq:SI:Dmu_DF_single}.
    }
    \label{fig:SI:single}
\end{figure*}

Each of the interfaces is tested individually while in their stacked configuration, such that the two contacting surfaces are the same as in the $n > 1$ configuration.
To test the interface $a$, we only keep spring $a$, while the slabs below it ($i < a$) are clamped such that their absolute position is fixed.
That way, only the bottom interface of slab $a$ is allowed to slide.
In this case, the slip $\Delta R_a$ and the macroscopic force drop are related as
\begin{equation}
    \Delta F = K \Delta R_a \, a h / H,
    \label{eq:SI:DR_DF_single}
\end{equation}
which is verified by the data in \cref{fig:SI:single:a}.
The drop of frictional resistance then reads
\begin{equation}
    \Delta \mu_a = \frac{K \Delta R_a}{N} = \frac{\Delta F}{N}\frac{H}{ah} .
    \label{eq:SI:Dmu_DF_single}
\end{equation}
In addition, to ensure testing the interface at the same shear rate $\dot\gamma = 10^{-4}\mathrm{s}^{-1}$, despite the different heights onto which they are attached to the lever, the shear rate is adapted for each layer such that $\dot\gamma_a = \dot\gamma / a$.
With $\Delta R_a = T_a a h \dot\gamma_a = T_a h \dot\gamma$, we have that $\Delta \mu_a = K T_a h\dot\gamma / N$.
\cref{fig:SI:single:b,fig:SI:single:c} show $P(\Delta F)$ and the extracted $P(\Delta \mu_a)$, respectively, for $a = 1, 2, \ldots, 5$.
The interfaces exhibit significant differences in the mean values of $\Delta \mu_a$, ranging from $0.15$ to $0.3$ (with the differences in mean values larger than the scatter of the different $P(\Delta \mu_a)$) even though the material, preparation protocol, and environmental conditions are the same for each of these interfaces/experiments.

\section{Slip distributions}
\label{sec:SI:distributions}

\begin{figure*}[htp]
    \centering
    \includegraphics[width=\textwidth]{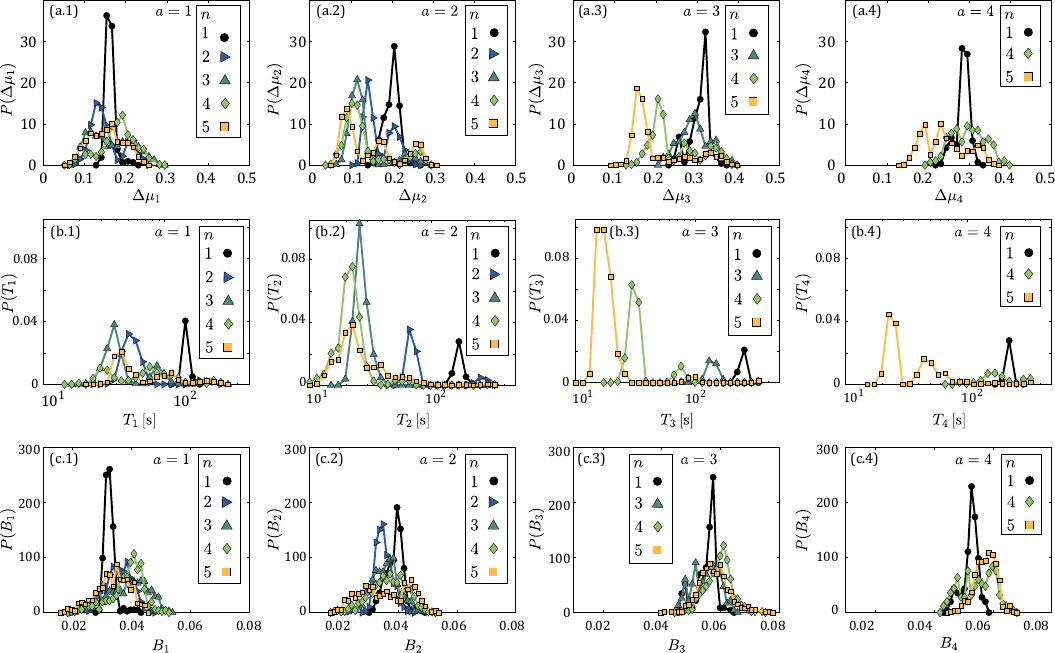}
    \caption{
        For each interface $a\in\{1..4\}$, probability distribution of the quantities associated to each slip event for all relevant number of active interface $n$.
        (a.1-4) Stick-slip amplitude $\Delta \mu_a$, (b.1-4) time between slip events $T_a$, and (c.1-4) effective aging rate $B_a$.
    }
    \label{fig:SI:B_distri}
\end{figure*}

For each detected slip event, we measure its stick-slip amplitude $\Delta \mu_a$ (\cref{eq:mu_a}), the waiting time since the last slip on the same interface $T_a$ and its effective aging rate $B_a = \Delta \mu_a / \ln(T_a / T_0)$.
In \cref{fig:SI:B_distri} we plot the corresponding probability distributions, $P(\Delta\mu_a)$, $P(T_a)$, and $P(B_a)$, for each layer $a$ and for different values of $n$ .

\section{Aging variations}
\label{sec:SI:aging_variations}

\begin{figure*}[htp]
    \centering
    \subfloat{\label{fig:SI:aging:a}}
    \subfloat{\label{fig:SI:aging:b}}
    \subfloat{\label{fig:SI:aging:c}}
    \subfloat{\label{fig:SI:aging:d}}
    \includegraphics{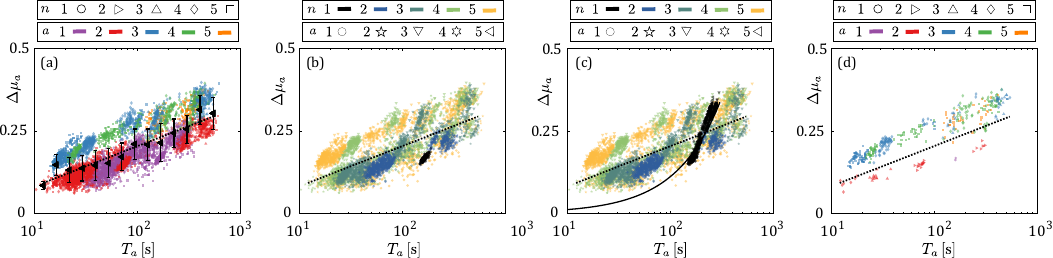}
    \caption{
        For each slip event, its stick-slip amplitude $\Delta \mu_a$ as a function of waiting time $T_a$, corresponding to the duration between the current slip event and the last slip event on the same interface.
        The dotted line in all the panels is the aging law $\Delta\mu_a = B\ln (T_a / T_0)$ (\cref{eq:aging}, dotted line) with $B = 0.053$.
        \protect\subref{fig:SI:aging:a}
        Reproduction of \cref{fig:5c}, where each event is colored by the interface $a$ it occurred, and its symbol reflects the number of interfaces $n$.
        The black markers are the average $\Delta\mu_a$ per bin of $T_a$, used to fit the aging law.
        \protect\subref{fig:SI:aging:b}
        The same data as previously are presented, with this time colors indicating $n$, and different symbols the sliding interface.
        \protect\subref{fig:SI:aging:c}
        On top of the previous data, the event measured on the interface sliding individually are included using black markers.
        Over this narrow range of $T_a$, disorder at the interface is sampled following the proportionality given by \cref{eq:disorder} (solid line).
        \protect\subref{fig:SI:aging:d}
        The data of (a) is biased to only events where there was no detected slow slip after the last slip event on the same interface.
    }
    \label{fig:SI:aging}
\end{figure*}

To provide a better visualization of the experimental measurements related to the aging of the interface, we display the corresponding data shown in \cref{fig:5c} and reproduced in \cref{fig:SI:aging:a} under various angles.
To emphasize the differences between the interfaces, \cref{fig:SI:aging:a} shows each slipping event colored corresponding to the interface $a$ where the slip occurred while the size of the stack $n$ is in marked by different symbols.
In contrast, to stress the role of adding layers to the stack, \cref{fig:SI:aging:b} shows the same data with a color code for $n$ and different symbols for each interface $a$.
In \cref{fig:SI:aging:c} we highlight the different roles of disorder and aging of the interfaces, both contributing to widening the measured distributions of $P(\Delta\mu_a)$.
On top of the data presented in \cref{fig:SI:aging:b} showing the logarithmic aging described by \cref{eq:aging} (dotted line), we included the events measured when the interfaces are tested individually (\cref{fig:5a}).
In this situation, each slip samples the distribution of interfacial strength caused by disorder, thus only a narrow window of $T_a$ is explored and $\Delta\mu_a$ is directly proportional to $T_a$ and follows a linear relation with time (\cref{eq:disorder}, solid line).
Finally, in \cref{fig:SI:aging:d} we display only the slip events where no slow slips occurred after the last slip event on the same interface.
To quantify the amount of slow slip between two events on an interface, we compare its relative position just after a slip event $R_a(t_n^+)$ and just before the next event $R_a(t_{n + 1}^-)$.
We consider that no slow slip occurred if $R_a(t_{n + 1}^-) - R_a(t_n^+) < 3\,\mu\mathrm{m}$.
The overall aging rate of the events without slow slip is not significantly different compared to the ones following some slow slip.
It thus seems that the amount of slow slip is not sufficient to rejuvenate the contacts and significantly affect the aging rate.

\section{Slow slip}
\label{sec:SI:slow}

\begin{figure*}[htp]
    \centering
    \subfloat{\label{fig:SI:smooth:a}}
    \subfloat{\label{fig:SI:smooth:b}}
    \subfloat{\label{fig:SI:smooth:c}}
    \includegraphics{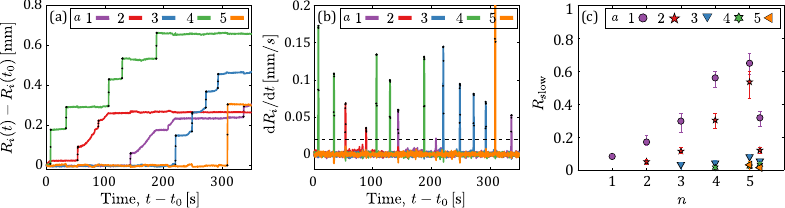}
    \caption{
        \protect\subref{fig:SI:smooth:a}
        Time series of the relative position $R_i(t)$ of the slabs for $n = 5$.
        The frames detected as belonging to a slip event are marked by a black dot.
        \protect\subref{fig:SI:smooth:b}
        Corresponding time series of the velocity of the slabs, after a running average over five frames has been applied.
        The horizontal dashed line marks the limit above which we consider that sliding is fast enough to be categorized as a slip event.
        \protect\subref{fig:SI:smooth:c}
        Same as \cref{fig:6b} with an additional measurement for $n = 5$ where the slabs are shuffled (${1,2,3,4,5}$ to ${4,3,2,1,5}$).
        The resulting data points are shown at $n \approx 5.3$ for visibility.
        After shuffling, it is still the lowest interfaces that shows most of the slow sliding.
    }
    \label{fig:SI:smooth}
\end{figure*}

To discriminate slip events from slow slip, we use the following velocity criterion.
The velocity of each slab is computed as $(R_i(t + \Delta t) - R_i(t)) / \Delta t$, with $\Delta t$ the time between frames ($\Delta t = 0.2\,\mathrm{s}$), and smoothed by a running average over five frames (\cref{fig:SI:smooth:b}).
Slip events are then defined by all clusters of at least three frames having a velocity greater than $0.02\,\mathrm{mm} / \mathrm{s}$ (identified by black dots in \cref{fig:SI:smooth:a} and \cref{fig:SI:smooth:b}).
In \cref{fig:SI:smooth:c} we verify that smooth sliding is consistently observed on the lowermost interfaces even if the slabs are shuffled.

\section{Numerical model}
\label{sec:SI:numerical}

\begin{figure}[htp]
    \centering
    \includegraphics[width=.45\textwidth]{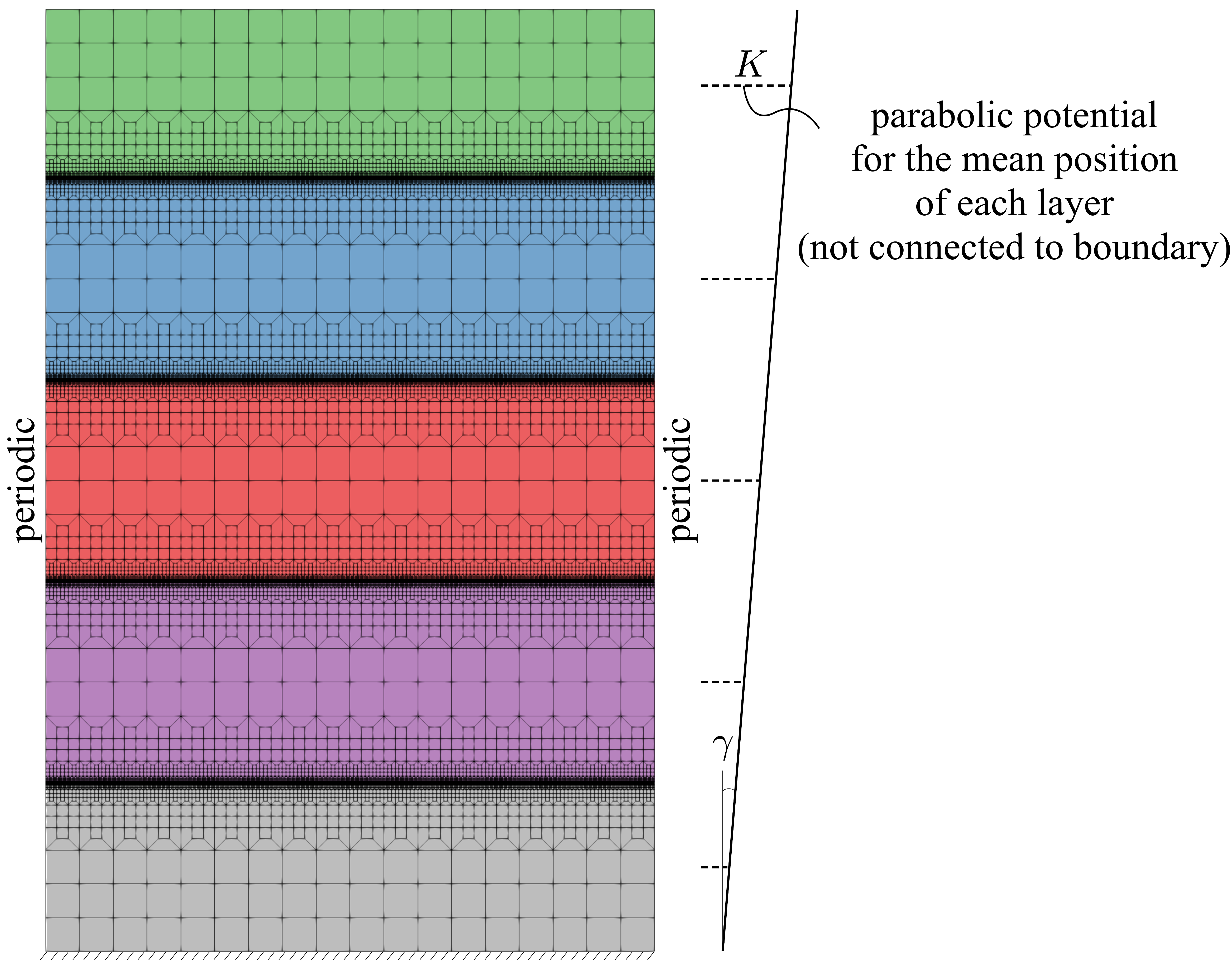}
    \caption{
        Finite element mesh of the numerical model for $n = 4$.
        The colors indicate the different layers (color coding is the same as in \cref{fig:1}).
        The positions of the bottommost nodes are fixed as indicated.
        The shear $\gamma$ is imposed by a parabolic potential of curvature $K$ associated to the mean position of each layer (the representation of the lever and springs is thus schematic).
    }
    \label{fig:SI:num:mesh}
\end{figure}

\textit{Parameters.}
All implementation details are identical to~\cite{deGeus2019}, and available in its supplementary material.
Furthermore, the entire implementation is open-source, see~\cite{FrictionQPotFEM} and its dependencies.
Compared to~\cite{deGeus2019} the only differences are: (1) The discretization, which now comprises $n$ frictional interfaces; see \cref{fig:SI:num:mesh}.
(2) The boundary conditions are such that the position of the bottom nodes is fixed while the top nodes are free.
The displacement of the right nodes is set equal to that of the left nodes corresponding to periodic boundary conditions in the horizontal direction.
Finally, all elements that constitute to an elastic layer $i = 1, 2, \ldots, n$ are connected to a parabolic potential of curvature $K$ that has its minimum at a shear $\gamma(\tilde t) i h$ (with $\gamma(\tilde t)$ applied using an event-driven protocol as described in the main text, such that $\tilde t$ only schematically represent time).
This potential is implemented as a force density and added to each node using the node's volume as a weight factor.
We use $K = 10^{-3}$ for stiff driving, and $K = 10^{-6}$ for compliant driving.
(3) We use a bulk modulus that is $4.5$ times larger than the shear modulus, in accordance with PMMA (see~\cite{deGeus2019} for the definitions of the bulk and shear modulus).

\textit{Units.}
Macroscopic stress $\Sigma$, defined as the equivalent deviatoric part of the macroscopic stress tensor $\Sigma_{ij} \equiv \int \sigma_{ij}(\vec{r}) d\vec{r}$ with $\sigma_{ij}(\vec{r})$ the deviatoric stress at a material point (i.e.\ at an integration point in our spatial discretization), and $\vec{r}$ the position.
Thereby $\Sigma^2 \equiv 2 \Sigma_{ij} \Sigma_{ij}$.
All reported stresses have been normalized by the typical yield stress.
For the numerical results, $\gamma$ is normalized by $\gamma_0$, which is defined as the shear corresponding to a yield event in a system in which all blocks of the first interface ($i = 1$) have a yield strain equal to the typical yield strain.

\section{Slip sequences -- numerics}
\label{sec:SI:sequences}

\begin{figure}[htp]
    \centering
    \subfloat{\label{fig:SI:num:Ta:a}}
    \subfloat{\label{fig:SI:num:Ta:b}}
    \includegraphics[width=0.45\textwidth]{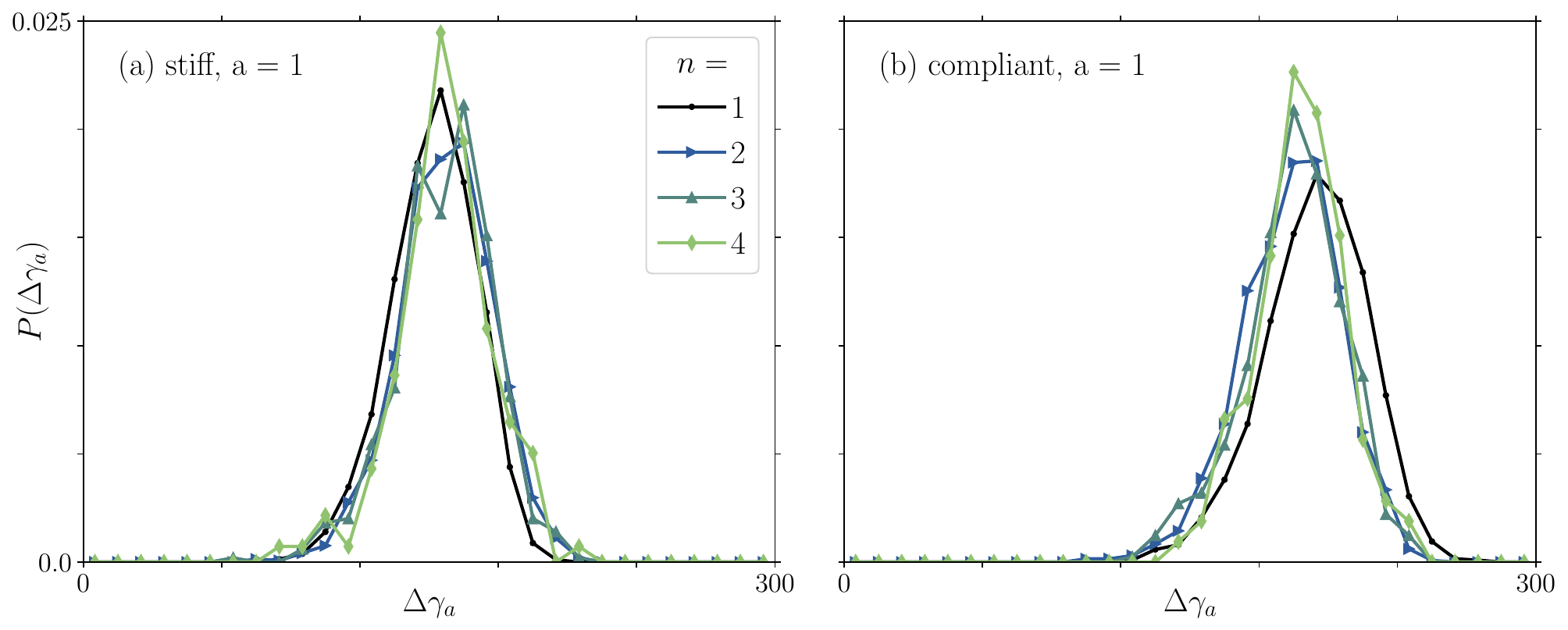}
    \caption{
        Shear increment $\Delta \gamma_a$ between slip events on the first interface ($a = 1$) as a function of $n$ for stiff driving \protect\subref{fig:SI:num:Ta:a} and compliant driving \protect\subref{fig:SI:num:Ta:b}.
        Note that we consider any slip event involving $a = 1$, i.e.\ also those in which multiple interfaces slip.
    }
    \label{fig:SI:num:Ta}
\end{figure}

We argue that the waiting time between slip on an interface, $T_a$, is a crucial quantity.
Since the numerics run a quasistatic protocol $T_a$ is not defined.
The closest equivalent quantity is the increment is the shear $\Delta \gamma_a$ between two slip events on a layer $a$.
We plot this quantity in \cref{fig:SI:num:Ta} for $a = 1$ for both stiff and compliant driving.
This plot should be qualitatively compared with \cref{fig:5b} of the main text.
However, only a coarse qualitative comparison is possible due to the following differences between the numerics and the experiments.
Firstly, the numerics follow a quasistatic loading protocol such that $\dot\gamma\equiv0$ and is independent of $n$.
Secondly, numerically we use a force density to control the mean position of each plate.
Consequently, the applied force \emph{does not} increase with $n$.
Lastly, the simulations are athermal and hence without aging.
What remains is that the higher fraction of single-slip events at lower $K$ leads to a slightly differently distribution $P(\Delta \gamma)$ for $n > 1$ in \cref{fig:SI:num:Ta:b}.
A better comparison with \cref{fig:5b} would require a lower $K$, currently outside our reach.
In addition the relationship between $T_a$ and $\Delta \gamma_a$ would have to be calibrated for the quasistatic protocol for each $n$.

\begin{figure}[htp]
    \centering
    \subfloat{\label{fig:SI:3a}}
    \subfloat{\label{fig:SI:3b}}
    \includegraphics[width=0.45\textwidth]{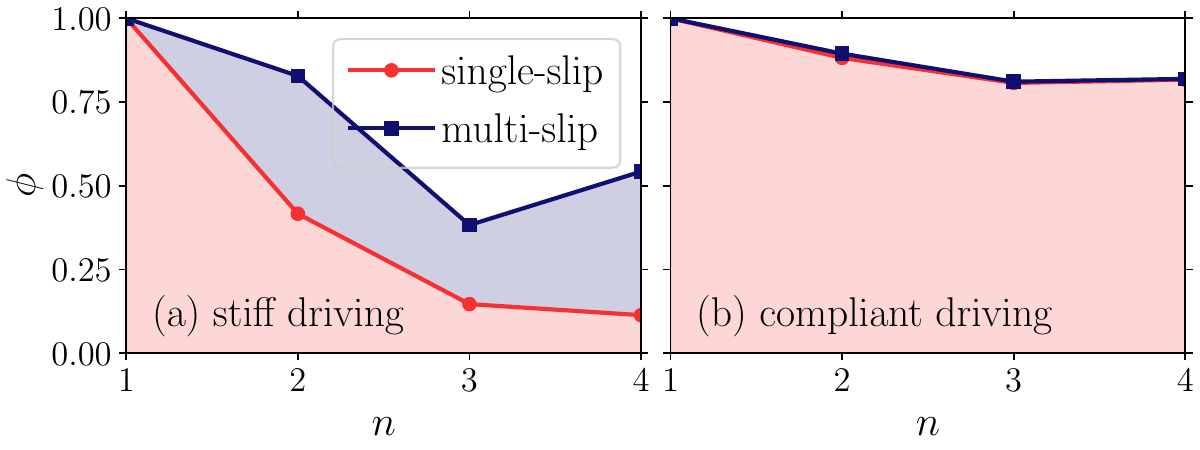}
    \caption{
        Fraction of slip events that are part of a sequence of single- or multi-slip events for (a) stiff and (b) compliant driving.
        The plot is constructed such that the blue line (square markers) corresponds to the total fraction of events that is part of a sequence of consecutive single- or multi-slip events.
        The red line (circular markers) is the fraction of events that are part of sequences of single-slip events.
        The area between the red and blue curves, highlighted in blue, thus corresponds to the fraction of events that are part of a sequence of multi-slip events.
    }
    \label{fig:SI:3}
\end{figure}

We comment on the fraction of events that are part of a sequence of single- or multi-slip events using \cref{fig:SI:3}.
For stiff driving, in \cref{fig:SI:3a}, sequences of single- and multi-slip events alternate.
As $n$ increases, such sequences alternate more often as reflected by the decreasing fraction of events that are part of any sequence (blue line).
Of these sequences, there is an equal partitioning in the number of single- and multi-slip sequences (the two highlighted regions have an approximately equal height for every $n$).
For compliant driving, the situation is simple: there are almost exclusively single slip events, see \cref{fig:SI:3b}.

\section{Front dynamics -- numerics}
\label{sec:SI:dynamic}

\begin{figure}[htp]
    \centering
    \subfloat[\label{fig:SI:event:a}]{\includegraphics[width=0.49\linewidth]{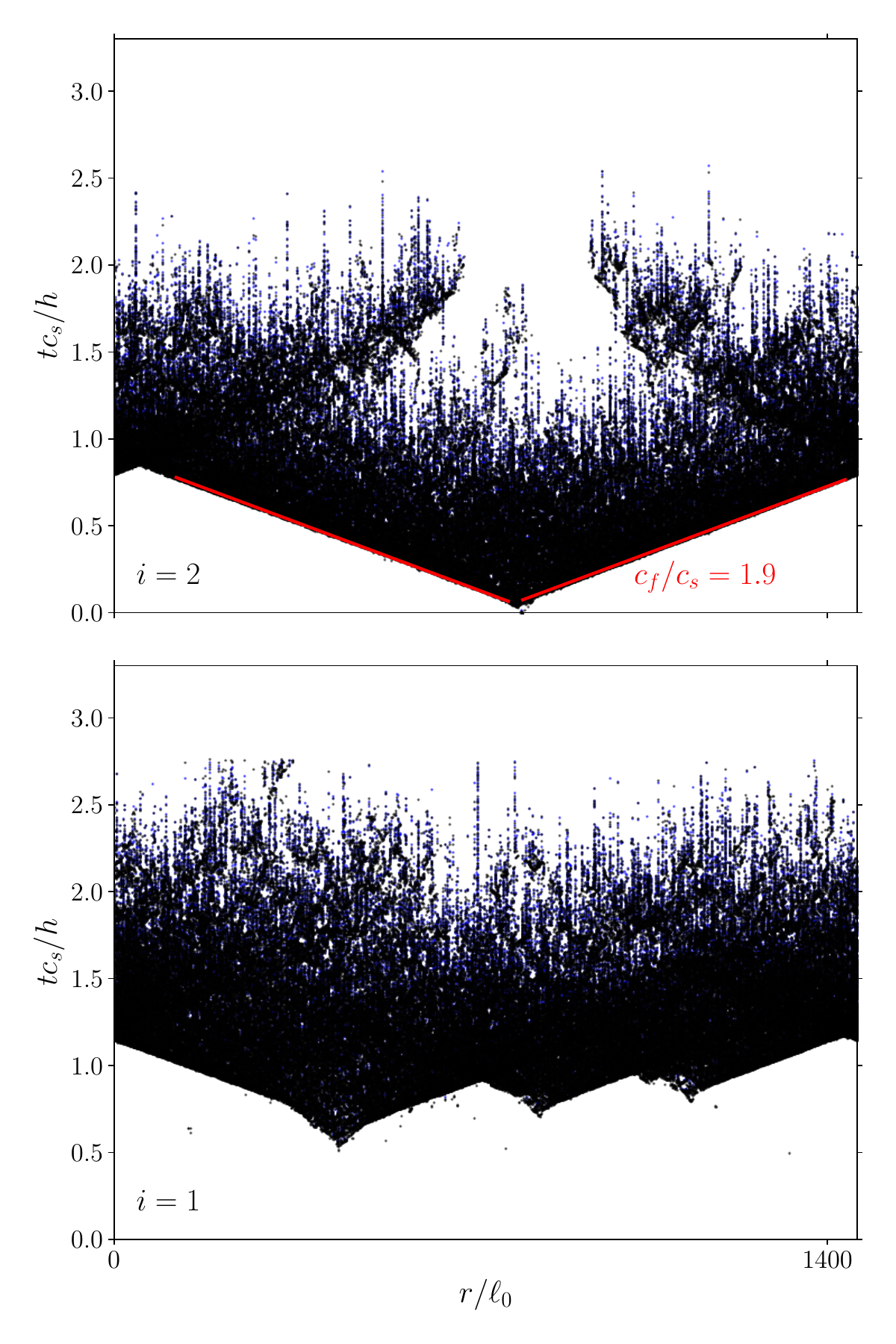}}
    \hfill
    \subfloat[\label{fig:SI:event:b}]{\includegraphics[width=0.49\linewidth]{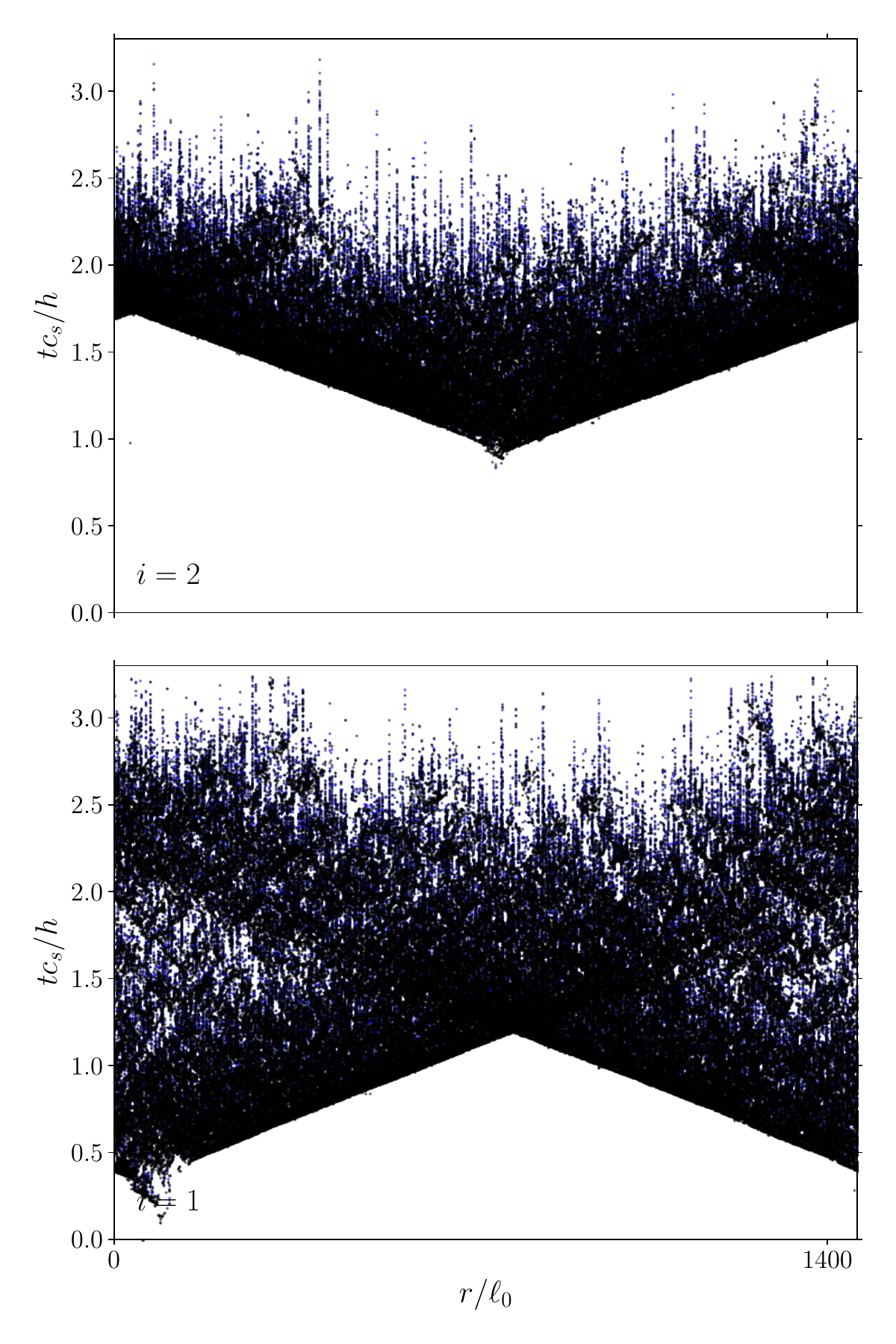}}
    \caption{
        (a, b) Event-map of two typical simulations for $n = 2$ .
        Each panel contains two plots above each other: one for each interface $i$.
        In each plot, the horizontal axis is the horizontal position $r$ normalized by the size of one block along each of the weak layers $\ell_0$.
        The vertical axis is the time $t$ since triggering the first failure anywhere in the system, normalized by the time it takes a shear wave to travel the height $h$ of each layer ($c_s / h$, with $c_s$ the shear wave speed).
        A point is drawn every time there is a failure (failure in the direction of applied shear in black, and in the opposite direction in blue).
    }
\end{figure}

\begin{figure}[htp]
    \centering
    \includegraphics[width=0.71\linewidth]{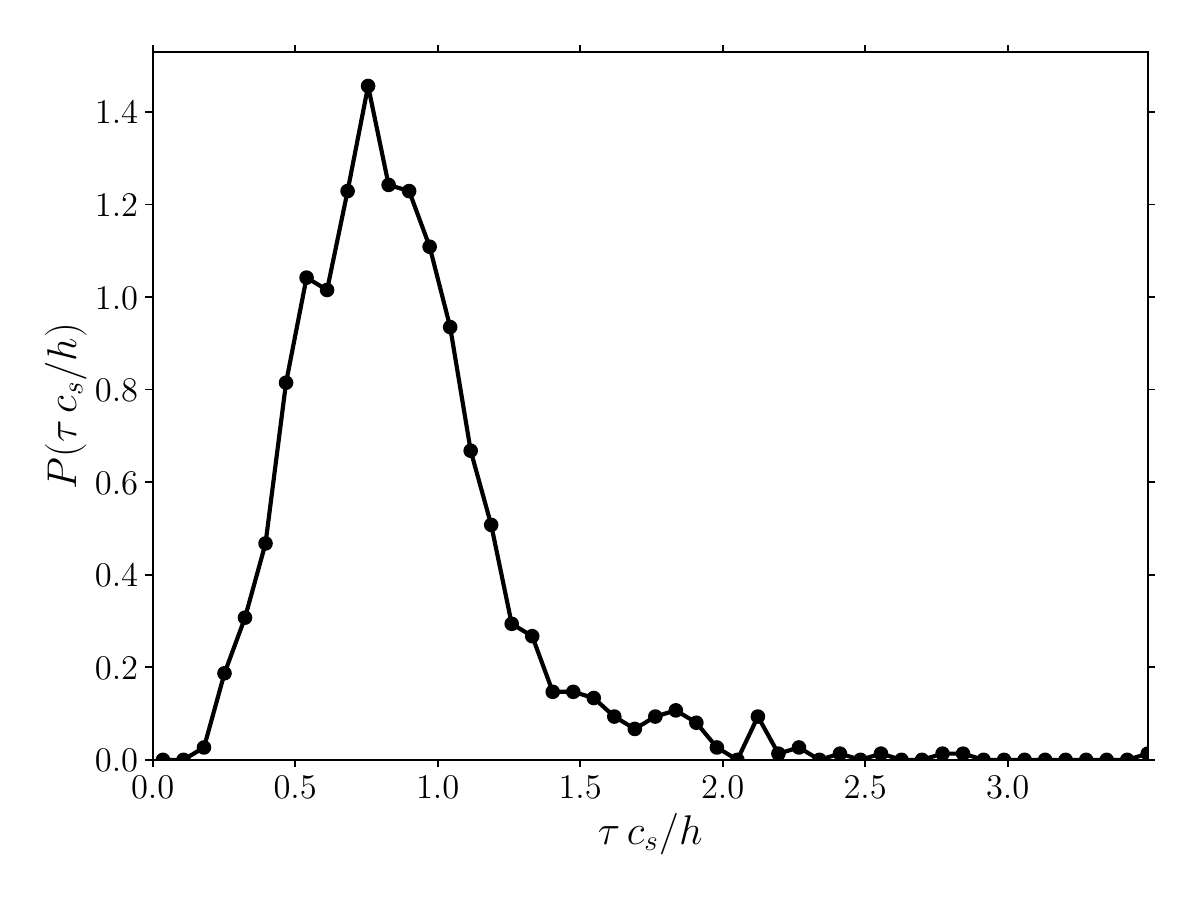}
    \caption{
        Probability distribution of the duration $\tau$ between activity on the primary and secondary slipping layer for $n = 2$, normalized by the time it takes a shear wave to travel one layer.
    }
    \label{fig:SI:duration}
\end{figure}

We briefly discuss the dynamics of a multi-slip event -- in which multiple layers slip during a macroscopic event.
For simplicity we consider a system with $n = 2$.
\cref{fig:SI:event:a,fig:SI:event:b} show event maps of two typical system-spanning events (in which all blocks of each interface fail at least once).
As observed, the applied infinitesimal increase of applied shear triggers an avalanche on one of the layers.
This avalanche propagates as a fractal object in space-time until it transitions to a ballistic event that propagates through the entire system \cite{deGeus2019,deGeus2022} with a front velocity $c_f$ of about twice the shear wave speed $c_s$ (see fit in \cref{fig:SI:event:a}).
With a delay of the order of the time a shear wave takes to propagate through the layer above or below ($c_s / h$, with $h$ the height of each layer), the other layer starts to fail.
In contrast to the first slipping layer, by chance avalanches can nucleate at multiple locations (e.g.~\cref{fig:SI:event:a}).
These avalanches start to propagate independently before linking-up.
We note that nucleating a single avalanche on the secondary slipping layer is also frequently observed (e.g.~\cref{fig:SI:event:b}).
We quantify the delay of activity on the secondary slipping layer.
We measure the duration $\tau$ between the first event on the first layer with activity and the first event on the other layer.
The distribution of $\tau$ is shown in \cref{fig:SI:duration}, confirming that the delay is of the order of the time it takes a shear wave to propagate through a layer.

\end{document}